\documentclass[acmsmall]{acmart}

\AtBeginDocument{%
  }

\setcopyright{rightsretained}
\acmJournal{PACMHCI}
\acmYear{2022} \acmVolume{6} \acmNumber{CHI PLAY} \acmArticle{223} \acmMonth{10} \acmPrice{}\acmDOI{10.1145/3549486}

\usepackage{graphicx}
\usepackage{array}
\usepackage{enumitem}
\usepackage{hyperref}
\usepackage{siunitx} % For typesetting scientific notation with \num{}
\usepackage{subcaption}
\usepackage{soul}
\usepackage{tabularx} % For table columns that fit the page width
\usepackage{csquotes} % For use of \enquote, easy quotation mark typesetting
\newcommand\ProjectName{Cine-AI}
\begin{document}
%\title{\ProjectName{}: Emulating Director Styles in Video Game Cutscenes}
\title[\ProjectName{}: Generating Video Game Cutscenes in the Style of Human Directors]{\ProjectName{}: Generating Video Game Cutscenes\\in the Style of Human Directors} %to stress procedural/generative aspect more


\author{Inan Evin}
%% \authornote{Both authors contributed equally to this research.}
\orcid{0002-9455-2097}
\affiliation{%
  \institution{Aalto University, Finland}
  \streetaddress{P.O. Box 11000 (Otakaari 1B) FI-00076 AALTO}
  \city{Espoo}
  \state{Uusima}
  \country{Finland}
  \postcode{FI-00076}
  }
  \email{inanevin@gmail.com}

\author{Perttu H{\"a}m{\"a}l{\"a}inen}
\orcid{0000-0001-7764-3459}
\affiliation{%
  \institution{Aalto University, Finland}
  \streetaddress{P.O. Box 11000 (Otakaari 1B) FI-00076 AALTO}
  \city{Espoo}
  \state{Uusima}
  \country{Finland}
  \postcode{FI-00076}
  }
  \email{perttu.hamalainen@aalto.fi}
  
\author{Christian Guckelsberger}
\orcid{0000-0003-1977-1887}
\affiliation{%
  \institution{Aalto University, Finland}
  \streetaddress{P.O. Box 11000 (Otakaari 1B) FI-00076 AALTO}
  \city{Espoo}
  \state{Uusima}
  \country{Finland}
  \postcode{FI-00076}
  }
  \email{christian.guckelsberger@aalto.fi}

\renewcommand{\shortauthors}{Evin et al.}

% A "teaser" figure, centered below the title and authors and above the body of the work.
\begin{teaserfigure}
  \centering
  \includegraphics[width=\textwidth]{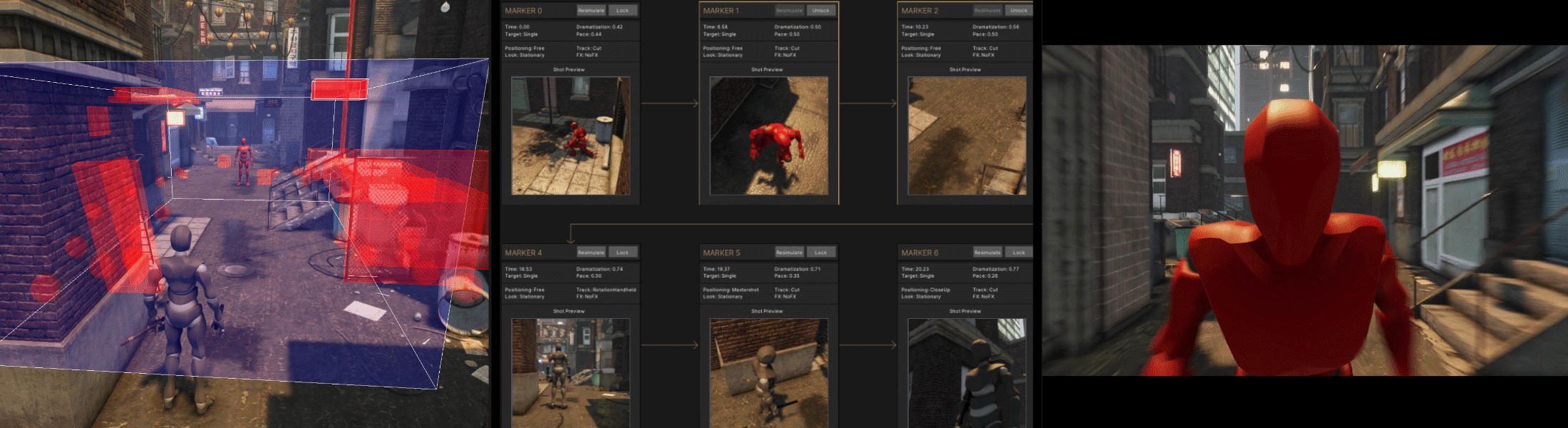}
  \caption{\ProjectName{} allows users to procedurally generate camera behaviour for in-game cutscenes by mimicking a target director's style. Left to right: cutscene edited by the user at design-time in the Unity game engine; Cine-AI storyboard interface allowing the user to fine-tune the generated scene composition; a still from the runtime cutscene in the style of the director Guy Ritchie.}
  \label{fig:teaser}
  %In clock-wise order: chase sequence from Guy Ritchie's \enquote{RocknRolla}, director style data resulting from annotating a selection of Ritchie's movies, Cine-AI storyboard interface to fine-tune cutscene generated in the director's style, still from the runtime cutscene.}\label{fig:teaser}
  %\caption{\ProjectName{} allows game developers to procedurally generate camera behaviour for in-game cut-scenes based on data from famous directors. It brings together camera motion, lens parameters and cinematography techniques in a storyboard which can be modified by the user at design-time. This enables the creation of high-quality game cutscene sequences with substantially less manual labor and cinematography knowledge, and without the costly acquisition of camera motion capture data. }\label{fig:teaser}
\end{teaserfigure}

%% abstract
% Must be no more than 150 words
\begin{abstract}
Cutscenes form an integral part of many video games, but their creation is costly, time-consuming, and requires skills that many game developers lack. While AI has been leveraged to semi-automate cutscene production, the results typically lack the internal consistency and uniformity in style that is characteristic of professional human directors. We overcome this shortcoming with Cine-AI, an open-source procedural cinematography toolset capable of generating in-game cutscenes in the style of eminent human directors. Implemented in the popular game engine Unity, Cine-AI features a novel timeline and storyboard interface for design-time manipulation, combined with runtime cinematography automation. Via two user studies, each employing quantitative and qualitative measures, we demonstrate that Cine-AI generates cutscenes that people correctly associate with a target director, while providing above-average usability. Our director imitation dataset is publicly available, and can be extended by users and film enthusiasts.
\end{abstract}

%% GET A CSS FROM HERE: https://dl.acm.org/ccs

%% CCS


%keywords

% TODO (guidelines: https://chi2022.acm.org/for-authors/presenting/papers/): 
% -Finish first revision pass
% -Video: We should have a section where we qualitatively characterize the results, basically replicating parts of what Inan says on the video
% -Get feedback from Elisa's group

\maketitle
\section{Introduction}

In-game cutscenes are non-interactive sequences in a video game that pause and break up gameplay. They are often used to progress the story, e.g. by showing critical events or conversations between the characters. Especially in high-quality productions, in-game cutscenes feature elaborate character animations, complex scene composition and extensive cinematography. Conveying a targeted experience requires deep engagement in digital cinematography and directing techniques \cite{ablan2002digital} to arrange the placement of potentially several virtual cameras, along with their motion and behaviours. Game companies thus may need to hire dedicated directors, cinematographers, and entire movie productions teams.

Alternatively, it is possible to leverage AI in co-creative tools \citep{kantosalo2016modes} to semi-automate the process of directing and camera management. This holds the promise of reducing the costs and repetitive labour associated with traditional cutscene production, and to empower game designers with the creative agency to realise their cutscene ideas themselves. Research has made considerable progress towards overcoming some of the major challenges involved, e.g. automated camera placement, subject visibility, shot continuity and scene composition \citep{christie, JhalaFirst, He, ChristiansonDeclarative}.
%However, cutscenes produced by existing systems %so far only manage to produce monotonous and repetitive scenes that 
%lack the sense of a human director's touch.
However, existing systems fail to produce compositions which feel internally consistent and uniform in style - features that can be particularly well observed in film cutscenes produced by eminent human directors. 

%In order to simulate the camera behaviour similar to a real director shooting the scene, there are numerous obstacles that a system needs to overcome. Problems such as camera placement, subject visibility, shot continuity and scene composition, along with creating a director feeling by imposing a unique shooting style have been partially addressed in a number of studies. For instance, \citet{christie} explains techniques and requirements for procedural camera control while taking cinematography techniques into account. The paradigm explained by the study can be used to create an autonomous system that is able to convey scene data with respect to cinematography rules. Another study relates to the problem for continuity and narrative construction in automated cinematography by presenting a discourse planning technique \cite{JhalaFirst}. Meanwhile some studies introduce broader systems that focus on creating full scene compositions, such as the model explained by \citet{He} or the declarative camera language proposed by \citet{ChristiansonDeclarative}. These studies present solid techniques and methodologies to overcome some or most of the problems, meanwhile failing to address the lack of a director's touch.

To address this shortcoming, we have developed and evaluated \ProjectName{}, a semi-automated cinematography toolset capable of generating in-game cutscenes in the style of a chosen, eminent human director, and according to user specifications. For our proof-of-concept, we analysed 160 movie clips of two directors with respect to different cinematography techniques, resulting in a style description dataset to be leveraged in procedural cutscene generation. %To support users in the creation of cutscenes, \ProjectName{} requires them to indicate significant points in their animation timelines, providing the basis for camera placement. 
%To facilitate intuitive design-time interaction, the system visualises the resulting scene composition in an interactive storyboard. 
%The user controls \ProjectName{}'s camera manipulation and output through intuitive design-time interaction with a timeline and storyboard. 
\ProjectName{} affords user control through design-time interaction with an animation timeline and a storyboard: Based on the user's selection of significant content on an animation timeline, the system visualises the resulting scene composition as an interactive storyboard, which the user can then tweak further. The camera placement, transitions and cinematography techniques in the generated result closely resemble a chosen director's style, providing uniformity and overcoming the stylistic shortcomings of previous work. We make the following contributions:

\begin{itemize}[noitemsep, topsep=4pt]
    \item A technique to co-create in-game cutscenes that mimic a target director's style in interaction with a user (Sec.~\ref{section:system}).
    \item \ProjectName{}, a novel, open-source cinematography toolset which realises this technique. To maximise accessibility and support immediate production use, we implemented the system in the popular game engine Unity (Sec.~\ref{section:system}).% and published it under an open-source license. \ProjectName{} is designed to rely on few computational resources, thus accommodating the requirements of games industry professionals (Sec.~\ref{section:system}). %The open-source code is available at \url{https://github.com/inanevin/Cine-AI} (Sec.~\ref{section:system}).
    %Accommodating the requirements of games industry professionals, the system has been developed to rely on few computational resources. 
    %Users of this methodology can create portable tools that are easily decoupled from any system and would be able to work with any director as long as one can provide sufficient input data in the means of cinematography techniques.    
    %\item An open-source implementation of the toolset in the popular game engine Unity 3D, fostering wide practical application and future research. The code is available at \url{https://github.com/inanevin/Cine-AI}.
    \item A publicly available, extendable style description dataset for use in \ProjectName{} or related projects. At present, it encodes the style of two eminent directors based on their usage of cinematography techniques (Sec.~\ref{section:dataset}).
    \item A user study that supports the system's ability to generate cutscenes in a target director's style (Sec.~\ref{section:study_comparison}).
    \item A usability study that suggests no major flaws (System Usability Scale grade B) and identifies topics for future improvements and research (Sec.~\ref{section:usability}).
\end{itemize}
Our implementation and dataset are available under the MIT license on GitHub (\url{https://github.com/inanevin/Cine-AI})% and through the Unity Asset Store
, allowing for the straight-forward evaluation, extension and application by researchers and end users. Study materials and results are provided as Supplementary Material. A Supplementary Video presents an overview of the system, its design-time user interaction, and its director imitation functionality.
%We set out by relating \ProjectName{} and our individual contributions to existing work in procedural cinematography, introducing core background concepts along the way.
% and the corresponding background concepts. %This is followed by an introduction to our dataset in Sec. , and the description of the \ProjectName{} system and its user interface components in Sec. . We present our two studies in Sec. and ,

%Our implementation is developed in Unity game engine that provides a toolset for animation, sequencing and timeline management as well as an Application Programming Interface (API) for implementing custom GUI tools. 

\section{Background And Related Work}
\label{sec:related_work}
%The amount of material in the field of simulating a director's style in the context of games is scarce. We hence relate to existing contributions to 
%While there exists no prior research on procedural cutscene generation for video games, some core challenges have been addressed in virtual 3D environments more generally. 
In each of the following subsections, we introduce one challenge of automatic cutscene generation, how it has been addressed in previous work, and how it is overcome by \ProjectName{}. 
%We hence relate to contributions to procedural cinematography in virtual 3D environments more generally, describing the specific challenges addressed in the following subsections. 
We moreover consider a small body of work on generating storyboards, a central means of interaction with \ProjectName{}. We only take into account existing publications on storyboards for design-time manipulation, but omit work dealing with e.g.~the creation of storyboards for Web video in-sights \cite{furini} or the use of storyboards in live-action movie production \cite{Shakil}, as it is out of scope. We also exclude work on physical cinematography (e.g.~\cite{xie:hal-01819103, rogerio}), since the proposed solution space for e.g.~subject visibility, camera orientation and occlusion differs too much from Cine-AI's 3D game engine environment.

%We introduce the corresponding background concepts alongside the related work where needed. %These studies revolve around the ideas of creating systems that are able to interpret virtual scene data and generate a meaningful output that can be used to create scene compositions, including camera placements and cinematography techniques. They mostly focus on overcoming individual problems such as the automation of camera placement, subject visibility, scene continuity and narrative progression. There is also some, albeit much more limited, research on creating storyboards in order to utilize these notions in a design-time based manner. Below, we review some of the previous studies briefly in the context of problems mentioned above and discuss their key points.

\subsection{Camera Placement}

\citet{Arijon} introduces idioms for appropriate camera positioning during dialogue sequences. He holds that, in a single character environment, the camera should be placed within the field-of-view cone of the subject. For multiple characters however, he suggests to create a line of action using the middle points of the characters' positions. \citet{He} exemplify a set of rules of thumb and constraints implemented in their \textit{Virtual Cinematographer} that tend to submit to broader approaches and idioms defined by filmmakers throughout the years. 

\ProjectName{} adopts \citeauthor{Arijon}'s \cite{Arijon} idioms and \citeauthor{He}'s \cite{He} rules of thumb for camera placement. It moreover uses common cinematography approaches to determine camera angles.

\subsection{Subject Visibility}
The challenge of camera placement is closely intertwined with that of subject visibility, as virtual objects, characters and particle effects in a cutscene can block a clear shot of the the target subject. \citet{lino:hal-00535865} propose to identify subject visibility areas for camera placement through \textit{visibility volumes}. This technique involves projecting the 3D environment onto a 2D image, which is then divided into smaller cells. By iterating over the pixel information on these cells, one can swiftly determine key areas of subject visibility. Meanwhile \citet{Oskam} propose an algorithm for visibility-aware path-planning in a virtual environment. It uses visibility data for various parts of the scene along with pre-computed representations of collision-free paths to execute a camera transition with clear subject focus during runtime. \citeauthor{rucks}' \cite{rucks} CamerAI uses reinforcement learning to optimise subject visibility. More specifically, they use a continuous flow of runtime collision information along with a custom reward function to train a neural network camera control policy through Proximal Policy Optimization. The resulting model is capable of successfully tracking target objects in unseen 3D environments similar to those in their training maps. \citet{burg} introduce a fully dynamic occlusion avoidance system to maximise target visibility. Drawing on previous contributions \cite{linoref,linoref2} on Toric coordinate systems, they project occluder information into Toric Space to create an occlusion map. This is then used to derive an anticipation map by looking up the velocity information of the occluder vertices. This anticipation map is fed into a system that calculates appropriate camera orientation in runtime.

Both \citeauthor{rucks}' \cite{rucks} as well as \citeauthor{burg}'s \cite{burg} approaches provide accurate subject visibility in runtime. Crucially though, \ProjectName{} is designed for use in game engines, allowing us to address subject visibility at design-time in order to free resources at game runtime. Consequently, \ProjectName{} adopts a similar approach to \citet{lino:hal-00535865}, who analyse the current scene from the camera's point of view and perform %world-based 
raycasting to calculate \emph{visibility volumes}. We employ a similar analysis and raycasting within \textit{scene proxies} -- user-defined areas in the scene -- to calculate potential collisions at design-time and ensure collision-free camera motion during runtime. These proxies are also inspired by \citet{Oskam} and will be elaborated on in our system description (Sec.~\ref{section:system}).

%% WE WROTE THIS BUT THEN REALIZED MORE PAPERS ARE SUBJECT VISIBILITY RELATED, LETS BETTER PHRASE BELOW
%%\ProjectName{} adopts a similar approach to \citet{lino:hal-00535865}, who analyse the current scene from the camera's point of view and perform raycasting to calculate \emph{visibility volumes}. \textcolor{red}{In comparison to \citet{rucks}'s work, \ProjectName{} achieves subject visibility by using both design-time and runtime calculations, emphasis being on the design-time stage. We use \textit{scene proxies} to ensure a collision-free camera motion which are inspired by \citet{Oskam} and will be introduced in our system description. The runtime visibility calculation stage of \ProjectName{} mostly plays a supporting role, by using a similar but simplified version of raycasting performed by \citet{rucks}'s work, in order to account for dynamic changes during a cutscene that were not present in the initial timeline. } 

\subsection{Shot Continuity}
Another challenge in a virtual environment is to create meaningful transitions, i.e. camera cuts and jumps that do not confuse the viewer but support the story, and achieve shot continuity. 
% A multitude of works address the issues of creating meaningful transitions and achieving shot continuity in a virtual environment. One method to achieve this is to abstract the whole animation timeline with various states, each state having a particular precondition or a goal to achieve \cite{JhalaFirst}. These can be anything in the context of the application, such as looking at a particular subject or precondition of some game event being triggered. Then the whole scene is treated as a one big state machine, generating shot sequences and motion plans to choose the best result according to the current state.
\ProjectName{} adopts the method by \citet{JhalaFirst}, who abstract the animation timeline into a network of states. Pre-conditions and goals facilitate transitions between these states, determining whether a state can be entered or left, respectively. This allows treating the scene as a \textit{state machine}, generating shot sequences and motion plans to choose the best result according to the current state. They propose a number of parameters including shot significance to rank and select the best generated sequences. Here, \textit{link planning systems} such as \textit{Longbow}, introduced by \citet{YoungDiscourse}, can be leveraged for sequence generation. 

Similar to Jhala and Young's \cite{JhalaFirst} use of user-parameters and shot ranking, \ProjectName{} provides various parameters that users can tweak to determine a shot's importance, pace and action value for ranking. In contrast to most other techniques though, the best possible shot is not auto-selected by the system at runtime, but it is offered to the user at design-time through storyboards. %\textcolor{red}{This allows us to avoid the computation of discourse planning links during the game, while providing complete flexibility to the scene designer.} 
This realises complete scene design flexibility. In addition, \ProjectName{} ensures scene and shot connectivity by relying on the cinematography rules derived from best cinematography practices; to be considered eligible, each generated shot undergoes a series of checks against these rules and is compared with the previous shots.

\subsection{Cinematography}
% How can camera placement, shot continuity and composition be brought together in a procedural fashion?
An umbrella term, cinematography denotes the art and technology of motion-picture photography \citep{BritannicaCinematography}. We aim to realise \textit{procedural cinematography}, focusing on how camera placement, shot continuity and composition can be brought together by algorithmic means. Tackling part of this challenge, \citet{ChristiansonDeclarative} propose a \textit{declarative camera control language} to formalise a selection of common cinematography idioms for automation. The formalisation of such idioms makes it easier to categorise rules of a scene along with the goals of particular shots. A similar representation is used by \citet{JhalaFirst} to declaratively represent storytelling plans. Along with selecting the best shot sequences, it becomes possible to create uniformity across transitions that meaningfully convey the narrative elements. Similarly, \citet{Karp1993:15} as well as \citet{Drucker1994IntelligentCC} propose encoding idioms and rules in \textit{film grammars} to generate a set of shot sequences via top-down analysis. Such sequences allow for the calculation of motion planning to achieve certain visual goals. Since this technique relies on animation timing information, it is suitable for realising static cutscenes, but does not allow for cutscenes to unfold dynamically based on the previous game state. % virtual scenes, but it could not accommodate dynamic changes in video games.
\citet{jiang2020example} propose an approach for translating cinematography information from video clips to runtime camera motion. It consists of extracting cinematography features from sample video clips, deriving camera behaviour from these features with a deep learning algorithm, and applying this behaviour to a 3D camera motion controller. While both projects are in the realm of automated cinematography, \ProjectName{} focuses on the complementary tasks of defining directorial styles and providing users with a rich interface to manipulate the generated cinematic shot composition.

Instead of solely focusing on film idioms, \ProjectName{} creates meaningful scene compositions based on real director data, derived from 160 different movie clips. We build on existing work by representing the extracted director style data as hierarchy of idioms, enabling our algorithm to determine the best cinematography technique based on the user's choice of director and scene parameters. \ProjectName{} allows for cutscenes to unfold dynamically based on the previous game state, rendering the production of different, static cutscenes for each possible gameplay outcome obsolete. 
%We use the data extracted from movie clips in an hierarchical manner, which then help our algorithm to determine the best cinematography technique to use depending on the user's choice of director and scene parameters. 
%By resting \ProjectName{} on techniques employed by famous directors, the generate shot sequences and transitions resonate meaningfully within the virtual composition.

A remaining challenge consists in selecting the best possible cinematography techniques and shot sequences for a particular time during a specific in-game cutscene. \citet{Lima} have addressed this with supervised machine learning (ML) methods. They use \textit{support vector machines} (SVMs) to determine the best possible shot selection for a specific cinematography technique, based on the scene type, number of actors and their features. Such an approach can produce great results, but offers little control to the user.% and typically comes with high computational demands. 

\ProjectName{} only leverages machine learning for the initial definition of the style description dataset. At runtime, it relies on rules and parameters that warrant transparency, resource-efficient execution, and user control. 

\subsection{Storyboards}
We are not aware of existing approaches leveraging storyboards in 3D cinematography. Consequently, we outline related work on alternative storyboard uses in videogames, and on similar uses in different creative domains. 
%Although there exists previous work on storyboard generation, or even the extension of storyboards in live-action movie production context \cite{Shakil}, studies on implementing storyboards for procedural 3D cinematography purposes are lacking.

In a different videogame application, \citet{pizzi2008automatic} visualise simulated gameplay solutions to a level in the form of storyboards to retain designer control in interactive storytelling. \ProjectName{} similarly adopts storyboards as an established and familiar tool from game design. In contrast to \citet{pizzi2008automatic} though, who visualise gameplay as a sequence of a character's actions, \ProjectName{} visualises the composition of a game cutscene as a sequence of cinematography techniques.

\citet{Baikadi} bring storyboards to the domain of writing. They propose Narrative Theatre, a creativity support tool which leverages storyboards for narrative visualization. Users enter their narratives into the system, which are then fed through a natural language processor and a narrative reasoning module. Similar to \ProjectName{}, Narrative Theatre converts the resulting data into a storyboard, allowing quick iteration times and instant editing. Related, \citet{Skorupski2009InteractiveSG} introduce Wide Ruled, a story authoring tool harnessing the power of a plan-based story generation model. Wide Ruled uses a hierarchical structure of story actions and plot fragments to achieve a particular author-goal. 
%Similar kind of structure can be used in Cine-AI's storyboard generation to allow users to select potential cinematographic stories, defined via a sequence of shot selections.} 

Only loosely related, \citet{remiStoryboard} propose a storyboard language that describes each shot in terms of a sentence, which can be used to build software systems that convert formal shot descriptions into visual storyboard panels. This provides a way to automate the storyboarding process as well as virtual directing. \ProjectName{} in contrast uses storyboards as a visualisation and editing tool to present and tweak a produced scene. Acting as a summary of the user-customised scene, this interface allows to regenerate one or multiple shots, and to adjust the generation parameters. To the best of our knowledge, no such storyboard interface exists.

\section{Director Style Description Dataset}
\label{section:dataset}
To imitate director styles in automatic cutscene generation, our toolset relies on data which encodes the style of a certain director (Supplementary Video, 00:42). We next describe the acquisition and analysis of a proof-of-concept dataset. To extend this dataset to other directors, one only needs to replicate the steps described in Sec.~\ref{section:data_annotation}.
%we have watched sets of movie clips from two designated directors and collected different categories of data with respect to the usage of cinematography techniques. Machine learning algorithms such as Principal Component Analysis were used to increase the data interpretability, allowing us to find out which subsets of techniques differ the most between target directors. The statistical analysis process yielded an input for the toolset we have built in a real-time 3D game engine.

\subsection{Selection of Directors}
\label{section:data_director_selection}
Our dataset captures the characteristic style of two eminent directors: Quentin Tarantino and Guy Ritchie. We chose these directors for three reasons: 
\begin{itemize}
    \item People commonly consider both directors to have well recognisable and unique shooting styles, making them good candidates for evaluating our system's ability to capture and emulate human-perceived style differences.
    \item Many cinematography techniques are dependent on post processing, audio and visual effects,~i.e. technical domains that we chose not to consider at this stage of our project. The selected directors' styles differ particularly in terms of camera management -- the present focus of our modelling work. 
    %Many cinematography techniques are dependent on post editing, audio and visual effects,~i.e. technical domains that we chose not to consider at this stage of our project. % allowing coverage within the scope of this project.
    %\item As the data analysis is done within the research group, these two directors are the ones that the group is most affiliated with.
    \item Both directors can be assumed to be well known beyond an expert audience due to the wide popularity of their films, allowing for our results to be evaluated by a general readership. 
\end{itemize}
%Knowledge of the following data processing steps enables third parties to extend our publicly available dataset to other directors' styles, allowing for \ProjectName{} to be employed in projects with different stylistic demands.

\subsection{Data Annotation \& Aggregate Statistics}
\label{section:data_annotation}
As basis for our dataset, we extracted 80 one-minute clips from the most highly rated movies of each director on IMDB\footnote{Internet Movie Database, \url{http://www.imdb.com/}}. For added representativity, we chose half of the clips to be action-heavy, and the other half to be strong on dialogue. 

\begin{table}[ht]
    \centering
    \begin{tabularx}{\columnwidth}{l X} 
      \textbf{Technique} & \textbf{Description}\\ \midrule  
      God's Eye Shot & The camera is placed right above the subject, capturing an overhead angle. \\ \midrule
      Close-up Shot &  A shot taken in close range, usually to show detail of the subject's face. \\ \midrule
      Master Shot & A distant shot of the entire scene and all characters, e.g. for dramatisation. \\ \midrule
      Pan Shot & The camera is placed directly to either side of the subject while moving only horizontally. \\ \midrule
      Medium Shot & The camera is placed so that the subject is visible waist-up (waist-shot). \\ \midrule
      Long Shot & The camera is placed far away from the subjects such that they appear as indistinct shapes. \\ \midrule
      Free Shot & Camera placement at any distance between close-up and long shot. \\ \midrule
      Close-up Zoom & Close-up shot combined with slow lens zoom towards the subject's face. \\ \midrule
       Quick Zoom & The camera quickly zooms towards the subject, usually for dramatisation of a reaction. \\ \midrule
      Dolly Zoom & The camera zooms towards the subject whilst moving further away from it, keeping the subject the same size in the frame, thus undermining normal background perception. \\ \midrule
       Stationary Tracking & The camera tracks the subject without changing location, only by rotation. \\ \midrule
      Handheld Tracking & The camera tracks the subject, but shakes as if moved by hand. \\ \midrule
      Steadycam Tracking & The camera tracks the subject with a gimbal lock, i.e. without rotational or locational noise. \\ \midrule
      Slow-Motion & The in-game time processing is slowed down, impacting every animation and effect. \\ \midrule
      Cut & Simplest transition, in that the camera only switches views. \\
    \end{tabularx}
    \vspace{0.3cm}
\caption{Annotated cinematography techniques, some descriptions drawing on Arijon's \enquote{Grammar of the Film Language} \citep{Arijon}. A video with examples of each technique can be found at \protect\hyperlink{https://youtu.be/egHGcp3zqks}{https://youtu.be/egHGcp3zqks}.}
\label{table:cinematography_techniques}
\vspace{-2em}
\end{table}

For each clip, we counted how often a specific cinematography technique was used. We selected 15 techniques in advance. based on their potential to be implemented via camera manipulation (Table \ref{table:cinematography_techniques}). Each clip was also assessed in terms of its dramatisation level and the scene's pace, encoded as scalars between 0 and 1. We understand the dramatisation of a scene as the emotional intensity of character reactions within. We noted a high pace if the scene unfolds quickly. In contrast, we assigned low values to scenes with a calm character dialogue. For each of the 160 clips, we thus obtained a 17 element vector with 15 integers encoding how often a certain technique has been observed, and two real numbers representing dramatisation and scene pace.

To capture a director's style, we calculated aggregate statistics on the per-clip data. We determined how often a director used a specific technique, counting over all clips in the sample. Moreover, we calculated the means of the dramatisation and pace values over all techniques used by a specific director. These values allow us to capture the relationship between a director's use of a cinematography technique and the type of the scene. 

%Thus, \ProjectName{} can use the target director's dramatization and pace thresholds to sort the cinematography techniques based on the dramatization and pace values provided by users for their own cutscenes.

%They were not picked for a single source, but are based on the first author's memory of a wide range of subject-specific documentaries, articles, and books.

%, Subsets and Component Analysis
\subsection{Evaluation of Cinematography Techniques' Discrimination Potential}
For \ProjectName{} to generate recognisable output for a given director, we need to ascertain that the selected cinematography techniques are distinguishing features of the directors. %To ascertain this discriminative power, 
We trained a logistic regression \cite[p.~205 ff.]{bishop2006} model to predict the director of each annotated video clip, using the annotated cinematography technique frequencies as the regression features. Logistic regression models the probability of a director as proportional to $\sigma(w^{T}x+b)$, where $\sigma$ is the logistic sigmoid function, $x$ is a vector of features, $w$ is a (transposed) vector of feature weights, and $b$ is an optional scalar parameter. The absolute value of a feature weight can be interpreted as the feature’s estimated importance in correctly predicting the director. If a feature is not at all predictive of a director, logistic regression will assign it a zero weight.

\begin{table}[ht]
\begin{tabular}{m{8.2em} m{1cm}} 
  \textbf{Technique} & \textbf{Weight}\\ \midrule  
  God's Eye Shot & 0.84 \\ 
  \midrule  
  Steadycam Tracking & 0.61 \\
  \midrule  
  Handheld Tracking & 0.59\\ 
  \midrule  
  Close-up Shot & 0.53\\ 
  \midrule  
  Quick Zoom & 0.48 \\
\end{tabular}
\enspace
\begin{tabular}{m{7.7em} m{1cm}} 
  \textbf{Technique} & \textbf{Weight}\\ \midrule  
  Dolly Zoom & 0.41 \\ 
  \midrule  
  Close-up Zoom & 0.39 \\
  \midrule  
  Slow-Motion & 0.34\\ 
  \midrule  
  Medium Shot & 0.31\\ 
  \midrule  
  Long Shot & 0.38 \\
\end{tabular}
\enspace
\begin{tabular}{m{7.7em} m{1cm}} 
  \textbf{Technique} & \textbf{Weight}\\ \midrule  
  Free Shot & 0.15 \\ 
  \midrule  
  Pan Shot & 0.13 \\
  \midrule  
  Master Shot & 0.07\\ 
  \midrule  
  Stationary Shot & 0.05\\ 
  \midrule  
  Cut & 0.02 \\
\end{tabular}\vspace{0.3cm}
\caption{Logistic regression weights for each feature (cinematography technique) in ascending order, representing their contribution to distinguishing our directors, Quentin Tarantino and Guy Ritchie.} 
\label{table:logreg}
\vspace{-2em}
\end{table}
The logistic regression model is able to predict the correct director with an accuracy of 83.75\%, demonstrating that the selected features, although focusing on camera management only, can characterise exemplary director's styles. The feature weights are shown in Table \ref{table:logreg} in descending order of their discrimination potential. The identified values confirm our observations from the initial movie clip annotation. A god's eye view for instance is more commonly used by Quentin Tarantino (33.7\% of our clips) than by Guy Ritchie (8.75\% of our clips). Ritchie also uses steadycam tracking more often (24.6\%) than Tarantino (12.4\%). %On the other hand, overall cut/transition frequencies do not differ significantly between the directors (786 occurrences for Ritchie, 895 for Tarantino). 
Despite some regression weights being closer to zero, we did not discard any techniques in implementing \ProjectName{}, as they might later prove useful in discriminating between other directors. 

We argue that selecting the sample clips differently would not yield strong variations in the inferred styles, as there exists strong style uniformity amongst different scenes shot by the same director -- one of our primary motivations to mimicking the style of eminent human directors. %This uniformity is most strongly expressed in terms of camera work.

\begin{table}[t!]
 \begin{tabularx}{\columnwidth}{m{2.5em} m{4.7em} X m{6.5em}} 
  \textbf{Order} & \textbf{Category} & \textbf{Techniques} & \textbf{Default}\\
  \midrule
  1 & Positioning & Close-up Shot, God's eye Shot, Master Shot, Medium Shot, Long Shot, Pan~Shot, Free Shot & Free Shot\\ 
  \midrule
  2 & Look & Quick Zoom, Dolly Zoom, Close-up Zoom, Stationary Shot & Stationary Shot \\
  \midrule
  3 & Tracking & Steadycam Tracking, Handheld Tracking, No Tracking & No Tracking \\ 
  \midrule
  4 & FX & Slow-Motion, No FX & No FX \\ 
\end{tabularx}\vspace{0.3cm}
\caption{Cinematography techniques, including a default choice, mapped to camera manipulation categories. The categories enable a sequential camera manipulation workflow, in which a technique from each category is chosen and executed in the given order.}
\vspace{-2em}
\label{table:teccats}
\end{table}

\subsection{Categorisation of Techniques}
\label{sec:data_categorisation}
The techniques used in the statistical analysis are realised through different means of camera manipulation. For instance, a close-up shot requires re-positioning the camera near the subject's face \cite{Mascelli}, and the quick zoom technique requires re-adjusting the camera lens' field of view towards the subject \cite{wheeler2005practical}. To guide \ProjectName{}'s implementation, we assigned each of our technique to four categories of camera manipulation, ordered based on their sequential dependencies. Each of these categories in Table \ref{table:teccats} contains a default technique to fall back to if \ProjectName{} can not determine a specific technique to use later on in the simulation.
%This categorisation allows us to derive a natural, sequential camera manipulation workflow for \ProjectName{}

The technique chosen from the first category, \textit{Positioning}, defines where the system initially places the camera within the 3D scene geometry. Once placed, the camera always orients towards the user-defined target subject. Whether the camera lens is manipulated after this orientation step, e.g. in form of a zoom, depends on the corresponding technique selected from the \textit{Look} category. In the next step, positioning and lens manipulation can be augmented by techniques in the \textit{Track} category, defining potential means of moving the camera between two short markers. The final \textit{FX} category can comprise any form of camera manipulation that does not fit into the earlier categories. Since our study does not focus on effect post-production, we only allow for a slow-motion effect, or no effects at all. 

We note that dolly zoom could also be considered in the Positioning category, in that the camera is supposed to be moving on a line parallel towards the target. The Positioning category here however is concerned with static placements, not changes in position. We consider dolly zoom under the Look category, as it hosts techniques which require dynamic changes such as zoom ins/outs, field of view changes and similar.

\section{\ProjectName{} System and User Interface}
\label{section:system}

\begin{figure}[t!]
  \centering
  \includegraphics[width=0.9\columnwidth]{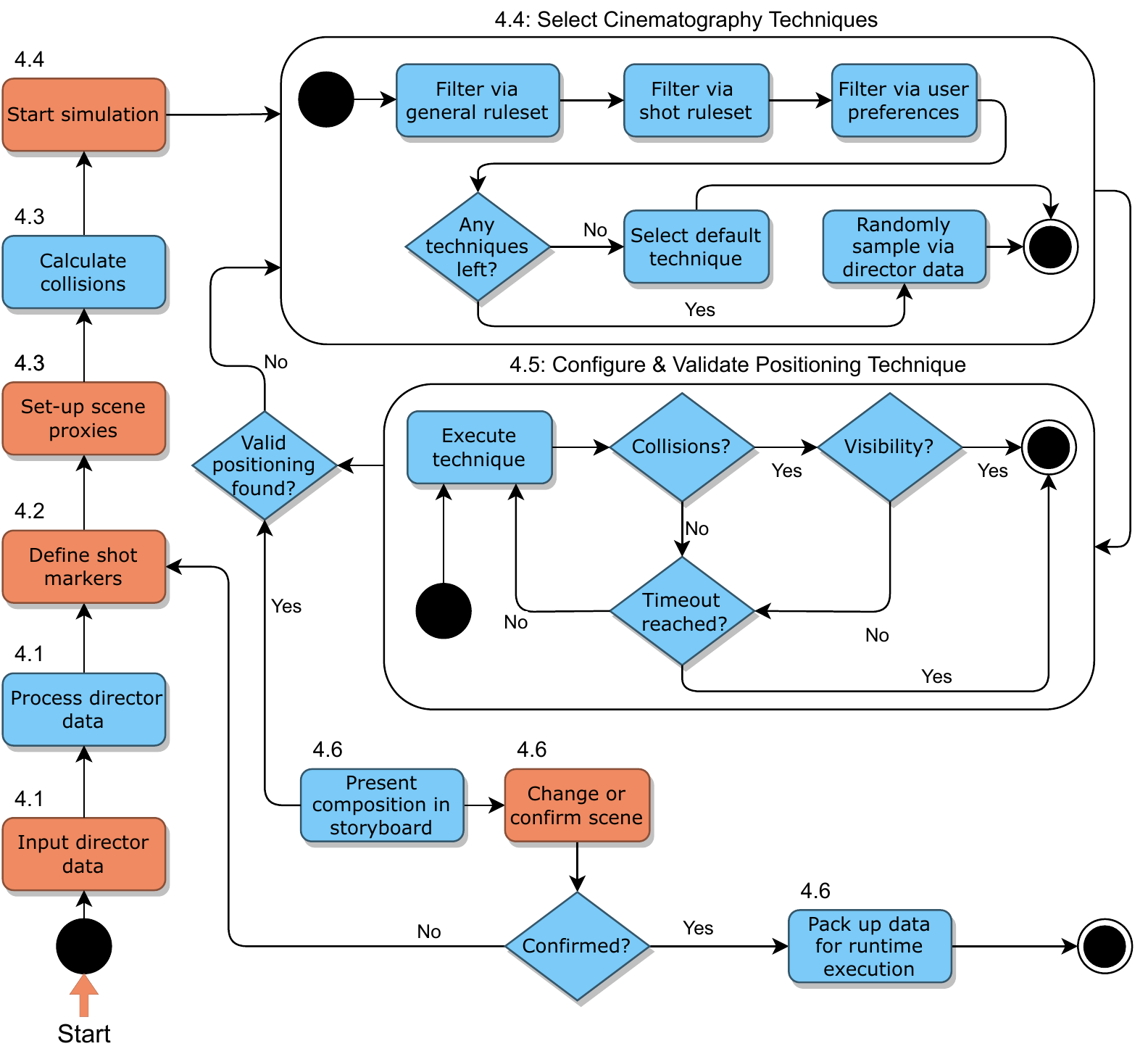}
  \caption{UML Activity diagram of \ProjectName{}'s design-time (sub-) processes (\textbf{rounded rectangles} - actions, \textbf{diamonds} - conditions, \textbf{circles} - start/end of overall or sub-process). \textbf{Orange} components require user interaction, \textbf{blue} ones are fully automated. Numbers refer to our subsections with detailed descriptions.}\label{fig:activityuml}
\end{figure}

We next describe the processes and user interfaces through which \ProjectName{} accomplishes the procedural generation of in-game cutscenes in a target director's style, with an emphasis on the design-time user interaction. The following subsections provide detail on one or multiple design-time (sub-)processes, as illustrated and referenced in Figure \ref{fig:activityuml}. In the last two subsections, we elaborate on the system's runtime support for dynamic in-game cutscenes, and summarise the user's required and optional interaction with the system.

In order to support \ProjectName{}'s straight-forward application in real-life game production, we have implemented the system in the popular Unity game engine. As an additional advantage, the Unity Editor provides an extensive set of tools for cutscene production, such as sequence and timeline editors, as well as immediate GUI libraries. 

%The interaction and processing steps of \ProjectName{} can be summarized as follows, with more implementation details and design rationale provided in the sections below.

\subsection{Director Data Input \& Processing}
\label{sec:system_data}

The user initialises \ProjectName{} by inputting a director imitation dataset, obtained through the process described in Sec.~\ref{section:data_annotation}. As a prerequisite to deciding which cinematography rules to abide by to generate camera behaviour in the style of the encoded director, \ProjectName{} calculates the conditional probabilities of each technique $t$ to occur within a category $C$ as
%In order to choose a cinematography technique for a shot marker in the timeline,
%, coded during the director analysis stage. When a director's data is imported,
\begin{equation}
p(t|C) = 
\begin{cases} f_t / \sum\limits_{t^\prime\in\ C}f_{t^\prime} &\text{, if } t\in C\\
0&\text{, otherwise.} 
\end{cases}
\end{equation}
This processing step relies on $f_t$, the total frequency of a technique observed over all clips, as comprised in the dataset.

\iffalse
\begin{figure}[ht]
  \centering
  \includegraphics[width=3in]{fig/directorData2.png}
  \caption{\textcolor{red}{Remove? }Categorized director data showing default (${P}_x$) for the cinematography techniques.}\label{fig:dirdata}
\end{figure}
\fi 

\subsection{Defining Shot Markers}
\begin{figure}[ht]
  \subcaptionbox{Shot markers in Unity Editor animation timeline\label{fig:markers_timeline}}{\includegraphics[height=1.6in]{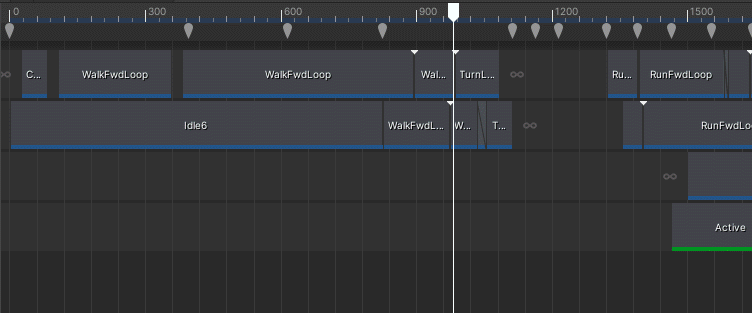} }\hfill%
  \subcaptionbox{Example marker settings\label{fig:markers_settings}}{\includegraphics[height=1.6in]{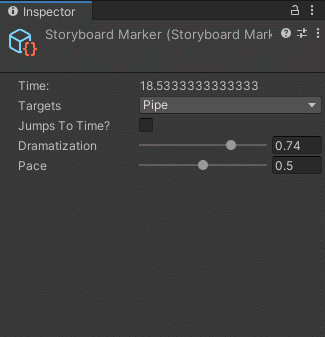}}%
  \vspace{-0.5em}
  \caption{Shot marker user interface. The user defines markers (\textbf{grey drop shapes}) in the Unity Editor animation timeline (a). The timeline comprises multiple tracks (\textbf{rows}) with animation and object activation nodes (\textbf{grey boxes with blue and green underlines, respectively}). For each marker (b), the user can set the camera target, and specify a desired transition dramatisation and pace.}
  \label{fig:markers}
   \vspace{-0.5em}
\end{figure}

The user defines shot markers in the Unity Editor animation timeline (Figure \ref{fig:markers_timeline}) to indicate at which time \ProjectName{} should insert cuts and transitions into the cutscene (Supplementary Video, 01:08). For each individual marker, the user can specify dramatisation and pace values, as well as which game object(s) the camera should focus on (Figure \ref{fig:markers_settings}) .
%By evaluating all markers,  \ProjectName{} will decide on a new cinematography technique per marker, thus a new camera framing, movement and transition, resulting in a cut. Each marker will correspond to a node in the generated storyboard, which will be further explained later.

\subsection{Scene Proxies and Collision Information}
\label{sec:system_proxies}
%\ProjectName{} collects static collision data from the cutscene by calculating scene proxies, 3D volumes that are used to define the boundaries of the cutscene.

\begin{figure}[ht]
  \subcaptionbox{Initial scene\label{fig:proxies_initial}}{\includegraphics[height=1.25in]{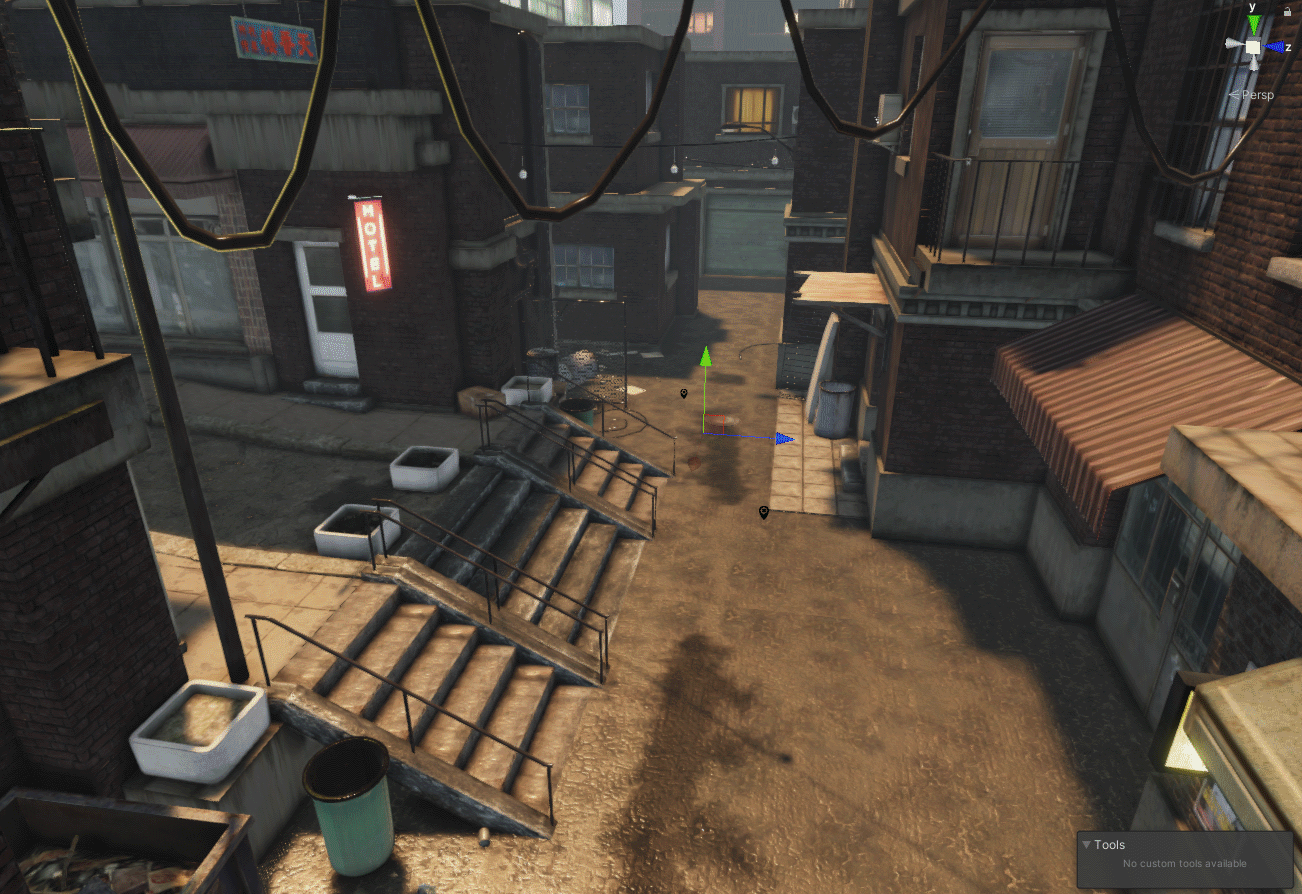}}\hfill%
  \subcaptionbox{User-defined scene proxy\label{fig:proxies_defined}}{\includegraphics[height=1.25in]{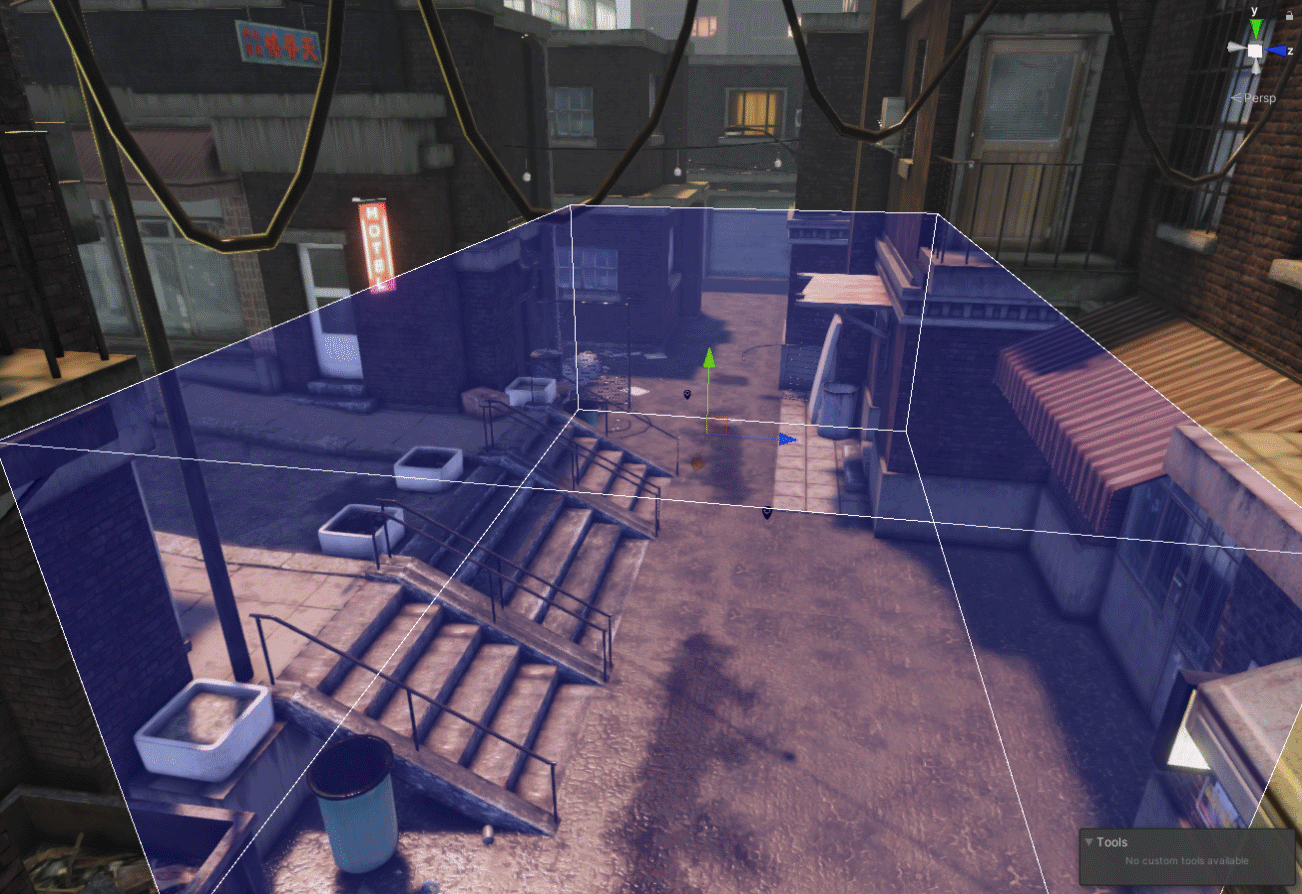}}\hfill%
  \subcaptionbox{Collision information in proxy\label{fig:proxies_evaluated}}{\includegraphics[height=1.25in]{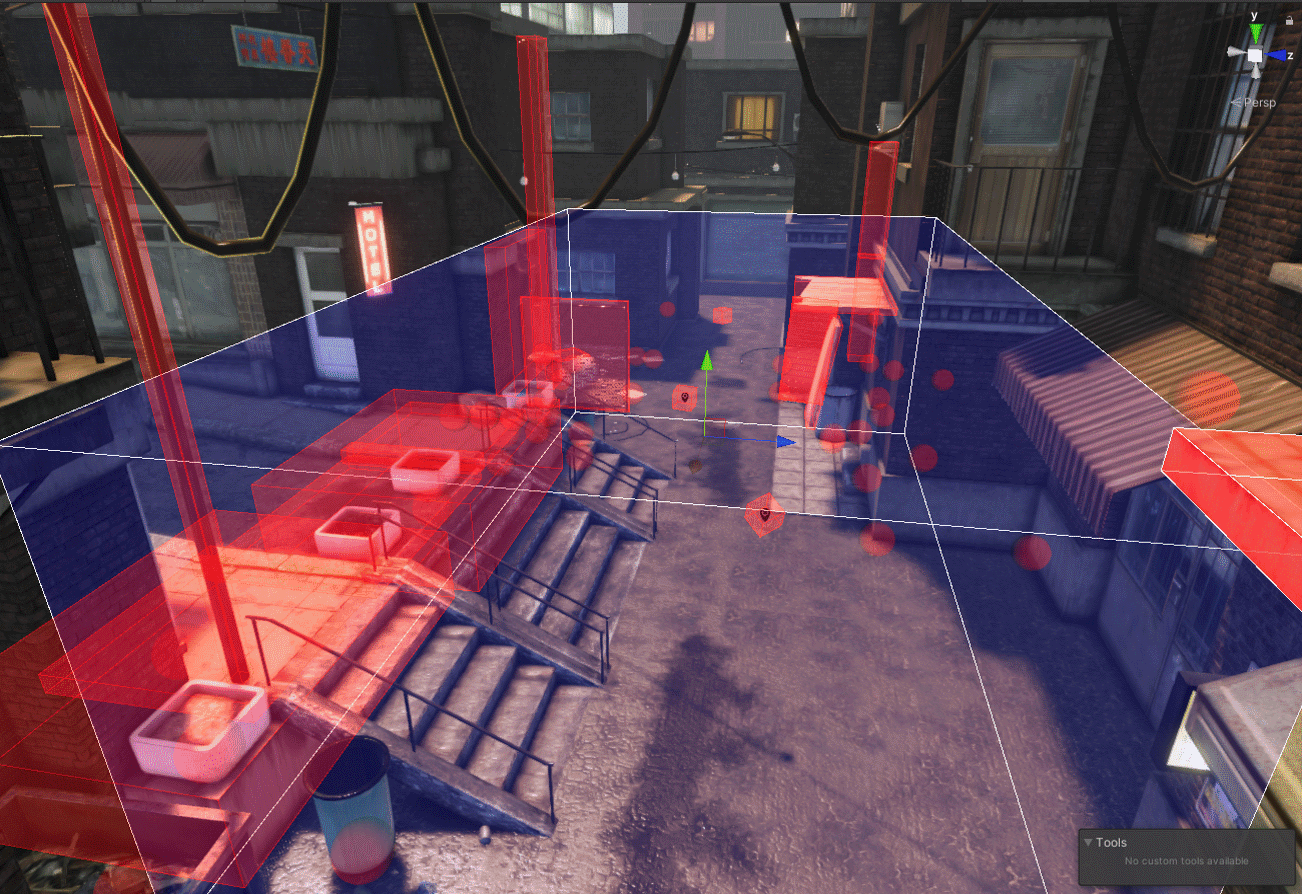}}%
  \vspace{-0.5em}
  \caption{Definition of scene proxies to obtain collision information. The user delimits in which area of the original game scene (a) the cutscene will unfold via a scene proxy (b, \textbf{blue box}). Cine-AI then calculates collision information (c, \textbf{red boxes}) within that area.}
  \label{fig:proxies}
  \vspace{-0.5em}
\end{figure}

To prevent camera clipping and realise collision-free camera paths, \ProjectName{} must obtain 3D collision information from the scene geometry. This calculation is expensive, but a cutscene is typically only set in a small part of the game world (Figure \ref{fig:proxies_initial}). For increased efficiency, \ProjectName{} requires the user to set up scene proxies, i.e. 3D volumes that delimit the area in which the cutscene takes place, and for which collision information is required (Supplementary Video, 01:28).
%\ProjectName{}'s interface provides parameters to adjust proxy settings per cutscene.

Once the user has set up the scene proxies (Figure \ref{fig:proxies_defined}), \ProjectName{} calculates and serialises the collision data at design-time, enabling efficient collision avoidance during runtime. This calculation is repeated for each marker in the animation timeline, respecting the change of posture, position or orientation of the characters and objects in the scene (Figure \ref{fig:proxies_evaluated}).
%To this end, \ProjectName{} goes through each shot marker, evaluating the cinematic sequence at the exact time of the marker so that the objects being animated will get in their respective poses for that time. Evaluation is followed by calculating the collision data for each object within the proxy, which is repeated for each marker in the timeline.

\iffalse
\begin{figure}[ht]
  \centering
  \includegraphics[width=0.8\columnwidth]{fig/proxyCalculation.png}
  \caption{Collision information (\textbf{red boxes}) calculated within the same scene proxy for two different markers, i.e. at two different times.}\label{fig:proxyCalc}
\end{figure}
\fi

%\ProjectName{} serializes the calculated collision data which can later be used while selecting camera positions as well as during the game while performing runtime collision avoidance.

\subsection{Selecting Cinematography Techniques}
\label{sec:system_selecting}

The user next initiates the simulation of all or specific shot markers (Supplementary Video, 01:40), yielding a set of cinematography techniques to be executed at the marker's time-step which (a) adhere to general cinematography principles, (b) respect the user's settings, and (c) imitate the target director. Here, we elaborate on the first sub-process in this simulation: the selection of suitable cinematography techniques (Table \ref{table:cinematography_techniques}) that satisfy these three requirements. To satisfy (a), the system filters the available techniques based on a general and shot-based ruleset, inspired by Kennedy and Mercer's Shotmaker system \citep{Kennedy}. The remaining techniques are then filtered further based on their match with the user-defined dramatisation and pace values (b). To imitate the style of the targer director (c), the system finally samples one technique per category via the conditional probabilities determined from the director dataset (Section \ref{sec:system_data}).  This process is repeated for each category (Positioning, Look, Tracking, FX;  Section~\ref{sec:data_categorisation}) in order, allowing to select e.g. a special effect based on the preceding tracking style.

%We use rejection sampling with technique sampling probabilities derived from the director style data and unsuitable techniques pruned based on hand-crafted rejection heuristics.

%based on rejection sampling. 

%A new camera behaviour, including the camera placement, angle, path, alignment and lens parameters is calculated based on the chosen techniques. 

%During the simulation, \ProjectName{} selects multiple cinematography techniques to use at the timestamp of the marker. We use rejection sampling with technique sampling probabilities derived from the director style data and unsuitable techniques pruned based on hand-crafted rejection heuristics. This procedure is performed consecutively for all categories, with the order as follows: positioning, look, tracking and fx. Following a specified order allows \ProjectName{} to consider the decisions given in the previous categories while choosing a cinematography technique for each category. This section explains the decision process and the reasoning behind it.

\subsubsection{General Rule Set Filtering}
%In order to implement both general and shot-based rule sets, we used a method inspired by the Shotmaker system of \citet{Kennedy}, where sets of rules direct the possible outcome cases, further eliminating the available cinematography techniques until only a handful is left. 
\ProjectName{} tries abides by a set of general cinematography rules, e.g.: %about cinematography in order to realistically compose scenes and process shot sequences. The said rules are considered as the ABCs of film-making. 
\begin{description}
\item[Triangle Configuration:] If multiple subjects are within view, the camera should be focusing the middle of a triangle determined by the position of each subject \citep{Arijon}. 
\item[Rule of Thirds:] To create a natural balance in the shot composition, the subject is placed on top of the cross-over points between imaginary lines \citep{heiderich2012cinematography}. The visibility of the subject is always prioritised over the rule of thirds.
\item[Leading Subjects:] In a continuously moving sequence, a tracking camera should come to rest before its target stops.
%\item{Line of Action:} An imaginary line of action connecting the major subjects in the scene shall be used during the calculation of camera positions.
\end{description}

As first step in the selection of a set of cinematography techniques, each technique from a specific category is checked against the general rule set and filtered out if in violation. For instance, if the 3D geometry at a particular time, defined by the marker, does not permit a long distance shot, then techniques like long shot or master shot are removed. %Likewise, if it is not possible to focus on multiple targets defined by a single marker at a time, all targets except the first one defined by the user are eliminated from the target list.

%\ProjectName{} tries to enforce this rule on target subjects, whilst allowing users to control this process through parameters
Due to the interactive nature of games and the complexity of their 3D scenes, exactly abiding by all rules is often infeasible. To increase flexibility and customisability, \ProjectName{} allows users to tweak, or \enquote{bend}, each rule through a set of individual parameters. For instance, users can weaken or intensify the rule of thirds by adjusting  minimum and maximum shot distances, obedience thresholds, and visibility checking options. %to affect how strictly \ProjectName{} should behave during the decision process.%as well as the visibility checking options can be used

\subsubsection{Shot-based Rule Set Filtering} \ProjectName{} checks each technique that adheres to the general rule set against another set of rules designed to facilitate good shot compositions and avoid unwanted repetitions. In this second filtering process, the system incorporates information about the techniques decided for previous categories and preceding markers. The shot-based rule set includes the following key restrictions:

\begin{itemize}
  \item It is not possible to use consecutive fast zoom techniques (quick zoom, dolly zoom).
  \item It is not possible to transit a master shot into a close-up shot. The camera should not be covering distances larger than a user-defined threshold at a single transition.
  \item It is not possible to use consecutive slow motion effects.
  \item It is not possible to apply any tracking technique that affects the camera position if dolly zoom is to be used in the Look category of the current marker.
\end{itemize}
%These rules are mostly derived from well-accepted rules of cinematography. Additionally, they allow \ProjectName{} to avoid unwanted repetitions. 

\subsubsection{User Preferences Filtering}
The resulting set is further refined by comparing the dramatisation and pace values of each remaining technique, determined from the director's data, to the desired values designated by the user in the shot marker settings. Any technique that does not fit the user requirements is eliminated. %If the selected technique does not fit the user's requirements, the sampling process is repeated until a suitable technique is found or a timeout is reached, in which case \ProjectName{} selects the default technique for the respective category (Table \ref{table:teccats}). 
The user can affect this process via threshold parameters, determining how strongly the marker's  dramatisation and pace values influence the selection. Based on these settings, the user can also skip the definition of dramatisation and pace values entirely. 

\subsubsection{Director Data Sampling} If the preceding filtering has eliminated all techniques, \ProjectName{} selects the default technique for the respective category (Table \ref{table:teccats}). If exactly one technique is left, it is selected and the sub-process stops. If multiple techniques are left, \ProjectName{} samples one technique at random based on the conditional probabilities calculated from the director data. This ensures that the selection yields techniques that are close to the target director's style, while affording stochastic variation if the user chooses to repeat the simulation.
%If not, the system proceeds to its last selections tep and 

%After the rule-based analysis of particular marker is done, \ProjectName{} is left with possible cinematography techniques to choose from. This is where the director data comes into play. A simple roulette selection function using the calculated probability of the techniques is performed in order to choose an available technique in the particular category.

\iffalse
\begin{figure}[ht]
  \centering
  \includegraphics[width=4.0in]{fig/image_2021-02-17_171824.png}
  \caption{ User interface showing user choices for decision function to use as well as dramatisation and pace thresholds. }\label{fig:userChoiceDistribution}
\end{figure}
\fi

\subsection{Executing, Configuring and Validating the Camera Positioning}
\label{sec:system_executing}
The second simulation sub-process consists of configuring, executing and validating the previously selected camera positioning technique at the respective marker. It yields settings for camera placement, angle, path, alignment and lens; the \textit{look}, \textit{tracking} and \textit{FX} techniques do not need further refinement, and are simulated at runtime. 

%, determining the camera placement, angle, path, alignment and lens parameters at the respective marker. 

%So far, \ProjectName{} would have selected a cinematography technique from each category and obtained the collision data from the scene geometry. 

%Next step for \ProjectName{} to perform is to find a suitable camera position based on the technique selected from \textit{positioning} category, and fix any \textit{look}, \textit{tracking} as well as \textit{fx} behavior to be simulated in runtime. 

\subsubsection{Execute Technique.} For each user-designated shot marker, \ProjectName{} firstly positions the camera according to the selected technique. For instance, the implementation of the close-up shot (Table \ref{table:cinematography_techniques}) dictates that the camera must be within a meter from the target, while the master shot requires it to be sufficiently far away to capture the whole scene. 

\subsubsection{Validate Collisions.} After the camera is placed, the collision data from the scene proxies (Section \ref{sec:system_proxies}) is used to determine whether the placement is valid or not. In the latter case, \ProjectName{} tries a new positioning based on the randomisation properties of the executed technique until deemed valid, or a timeout is reached. 

\subsubsection{Validate Visibility.} As a last step, the camera is oriented towards the subject point to perform visibility checks with a raycasting algorithm: a virtual capsule travels from the camera position towards the target position to check if there are any objects blocking the path in between. If such an occluder is detected, \ProjectName{} falls back to finding a new camera position and restarts the cycle. In case of a timeout, \ProjectName{} returns to the technique selection sub-process.\\

%The execution cycle is only performed for the \textit{positioning} category. Due to the dynamic nature of the games it is possible to have a multitude of animation and sequence possibilities during a cutscene. Thus, continuous camera behaviors such as a quick zoom or handheld tracking techniques are executed in runtime. 

While our implementations of the cinematography techniques in Table \ref{table:cinematography_techniques} follow the most common cinematography rules \cite{Arijon}, we anticipate projects in which these rules yield undesirable outcomes. For instance, \ProjectName{} implements Mascelli's \citep{Mascelli} recommendation for close-up shots to depict the subject from the chest to above the head. However, this might not be fitting for non-humanoid characters. To provide more flexibility to its users, \ProjectName{} allows to customise each cinematography technique through a set of individual parameters. 
%. For instance, \citet{Mascelli} explains that a close-up shot depicts the subject from chest to above the head, which our system tries to achieve for all close-up shots. However, 

\iffalse
\begin{figure}[ht]
  \centering
  \includegraphics[width=2.8in]{fig/techniqueOptions.png}
  \caption{Example of per technique parameters exposed for users.}\label{fig:techniqueSettings}
\end{figure}
\fi 

\subsection{Storyboard}

%After all shot markers are simulated (Section \ref{sec:system_selecting} and \ref{sec:system_executing}), the outcome for each marker is presented as one node in the storyboard, a preview of the calculated shot

\begin{figure}[ht]
  \subcaptionbox{Storyboard overview\label{fig:storyboard_overview}}{\includegraphics[height=2.29in]{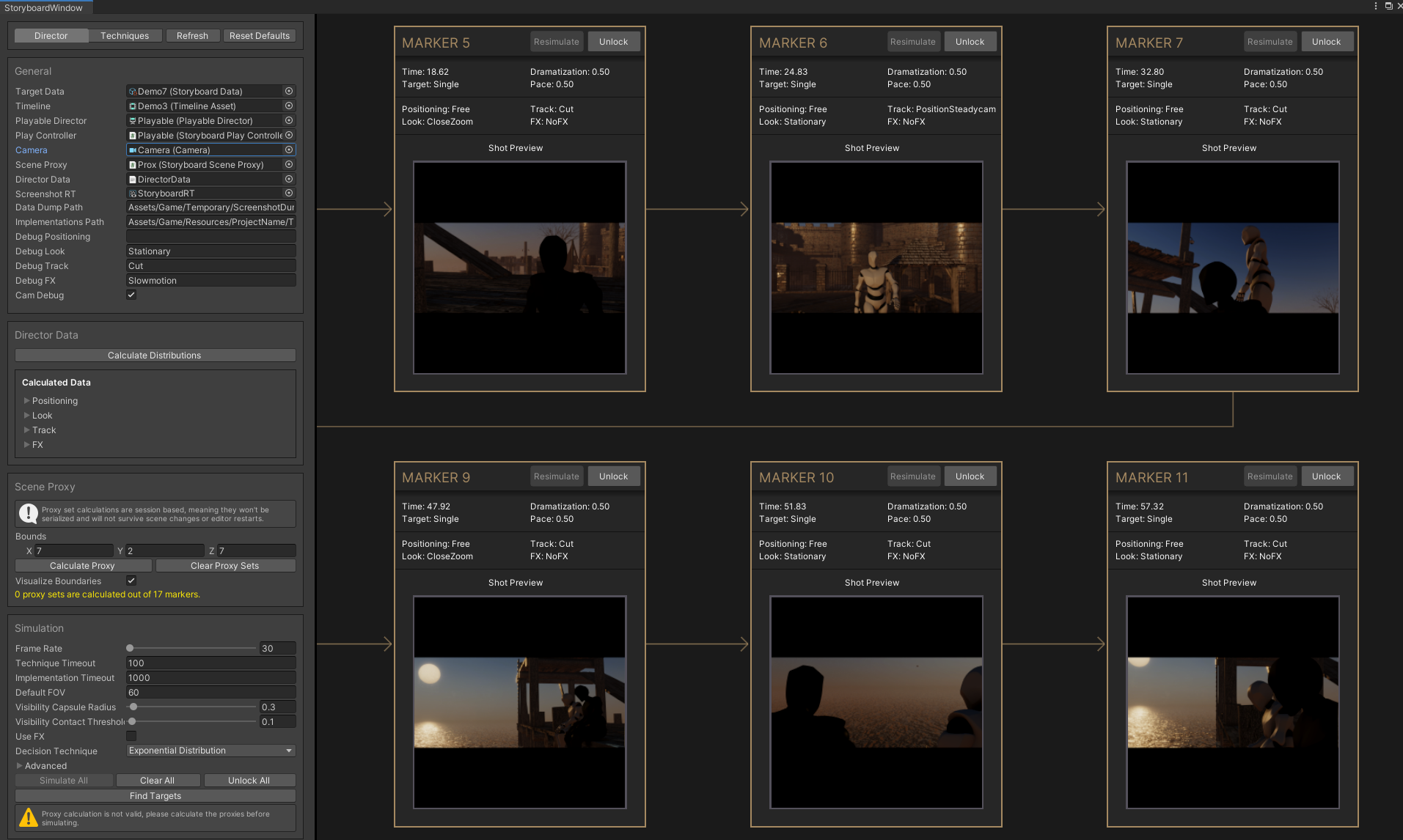}}\hfill%
  \subcaptionbox{Example storyboard node\label{fig:storyboard_node}}{\includegraphics[height=2.29in]{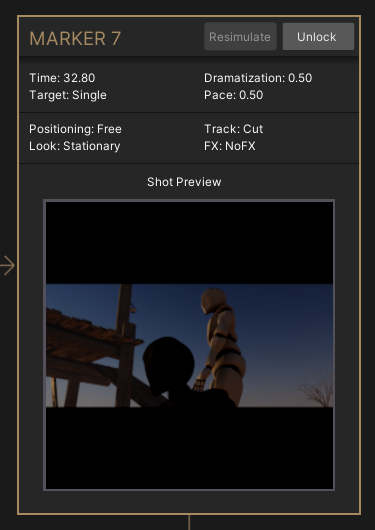}}%

  \caption{The storyboard user interface (a) comprises a vertical property window for editing simulation parameters on the left, and a node-based overview of each simulated shot marker on the right. The close-up view of a node (b) highlights a shot marker's parameters, selected cinematography techniques and shot preview.}
  %%\caption{Storyboard user interface. (a) Overview with each simulated shot marker, and (b) close-up view of a shot marker's parameters, selected cinematography techniques and a preview of the marker}
\end{figure}

Motivated by the prevalence of storyboarding in movie pre-production, \ProjectName{} provides the users with a storyboard interface as the central control element to review and tweak the scene composition at design-time, without running the game (Supplementary Video, 01:45). \ProjectName{}'s storyboard visualises the composition (Section \ref{sec:system_selecting} and \ref{sec:system_executing}) by displaying each shot marker as an individual node in sequential order (Figure \ref{fig:storyboard_overview}), comprising various setting and a preview of the selected camera angles (Figure \ref{fig:storyboard_node}). Moreover, the storyboard summarises information on, and affords control of the imported director data, scene proxy settings and simulation parameters (Figure \ref{fig:storyboard_overview}, left panel). After tweaking e.g. the individual techniques and cinematography rules, users can re-simulate the results for one or more markers. Each marker has a lock option to protect it from changes while re-simulating. 

Once the user is satisfied and confirms the cutscene composition, \ProjectName{} serialises all shot data in Unity asset files for use at runtime. It comprises e.g. target subjects, locations, the selected cinematography techniques with custom settings, and the simulated camera positioning. Our data representation is derived from the FILM language \citep{Charles}, a processing language designed to characterise the behaviour of camera motion within any given 3D world geometry. %Having done most of the heavy work done at design-time, \ProjectName{} can process the camera behaviour at runtime with very low resource requirements. 

\subsection{Runtime Execution}

The storyboarding completes \ProjectName{}'s design-time workflow; to play the in-game cutscene at runtime, the system listens to events fired as each shot marker is hit over time, and executes the corresponding cinematography techniques (Supplementary Video, 02:30). Although almost all required data is calculated at design-time, the system must still perform some operations at runtime to realise a crucial feature: its applicability to dynamic cutscenes. In such scenes, the behaviour of e.g. objects, animations and non-player characters (NPCs) can vary based on the preceding, player-induced game state; an NPC for instance might walk to different places in the scene, depending on the player's previous choices. 

Pre-calculating animations that reach across multiple frames at design-time is only feasible for static cutscenes. This applies to all techniques from the \textit{look}, \textit{tracking} and \textit{FX} categories, and \ProjectName{} consequently executes those at runtime. The potentially dynamic nature of the cutscene also implies that the 3D geometry might change in ways unanticipated during the design-time collision detection via scene proxies, thus necessitating runtime collision avoidance. If the camera penetrates an object, \ProjectName{} performs Haigh-Hutchinson’s algorithm \citep{Haigh} to move the camera along the object's surface. %whenever the camera detects a collision during the cutscene, it will slide on the surface of the collided mesh with an angle perpendicular to the surface's normal.
These runtime calculations require some additional computational resources, but enable the user to create only a single cutscene composition through \ProjectName{}, which can then adapt flexibly to the game at runtime.

\subsection{Summary of User Interaction}
We briefly summarise which steps (in \textbf{bold}) the user has to or can optionally perform in the interaction with \ProjectName{}, in which order, and to which end. As prerequisites, the user must perform the following steps in any order:
\begin{description}
    \item[Selection of Director Dataset (4.1):] The user must select a director dataset file, which will enable \ProjectName{} to imitate the director's style by calculating the corresponding thresholds during the procedural shot generation.
    \item[Shot Marker Definition (4.2):] Users is expected to have their cutscene animations ready on the timeline and define transition points within by means of shot markers. This is required for \ProjectName{} to parse the timeline and generate a shot sequence per marker, which will then be combined in the sequential storyboard representation.
    \item[Scene Proxy Setup (4.3):] The user must create scene proxy objects and setup their boundaries. This is required for \ProjectName{} to collect collision information from the scene pre-generation.
\end{description}
When everything is ready, users must \textbf{hit the generate button (4.4)} to create a shot sequence based on the markers, collision data, director dataset and parameters, which will be presented on the storyboard. Users can \textbf{re-generate the whole storyboard or just a collection of shots} if they desire. Users can \textbf{fine-tune technique-specific parameters} such as distance limitations, field-of-view limits and movement acceleration. They can moreover \textbf{adjust global parameters} such as timeout thresholds to tune \ProjectName{}'s execution speed on specific hardware. The shot sequence data is serialized immediately upon generation, and users can \textbf{switch into play-mode} to see their results in runtime.
%We motivate this design decision from two angles.
% to better respond to dynamic changes in the game scene.
%, e.g. player-induced variations in the game-state immediately before the cutscene.

%here, we briefly explain the system's runtime processing for the in-game cutscene. 

%As the users' cutscenes play, \ProjectName{} listens to events fired when each time a shot marker is hit, executing the cinematography techniques during the marker's timestamp. However, even though almost all the data related to the camera motion is calculated in design-time, performing some of the operations in runtime is still necessary to allow better responses towards dynamic changes in the scene. This section explains the runtime aspect of \ProjectName{}.

%For instance, the dolly zoom technique (Table \ref{table:cinematography_techniques}) requires moving the camera away from the subject while simultaneously decreasing the camera's field of view. However, 

\iffalse
\begin{figure}[H]
  \centering
  \includegraphics[width=4.0in]{fig/colavod.png}

  \caption{\textcolor{red}{Remove or replace with more illustrative image? }Camera performing line-casting for collision avoidance, detecting a wall behind and rejecting to move further towards back.}\label{fig:scriptableObject}
\end{figure}
\fi
\section{Experiment 1: Director Style Imitation}
\label{section:study_comparison}

In our first experiment, we probe \ProjectName{}'s ability to generate cutscenes that people can correctly associate with a specific target director. In a single-session, within-subjects study conducted through videoconferencing and online forms, we asked participants to associate cutscenes generated by \ProjectName{} with either director, Guy Ritchie or Quentin Tarantino. We then assessed quantitatively (a) whether the generated cutscenes for one director were consistently perceived as different from the other's, and (b) whether they were associated with the correct director. We complement these insights with qualitative information on our participants' decision-making process.

\subsection{Materials}
We produced eight different cutscenes with \ProjectName{}, half of which were shot based on Tarantino's and Ritchie's input data, respectively. Each cutscene is based on a different game scene. In order to adequately represent the defining features of our directors' styles, we chose cutscenes with action-rich content such as shooting, chasing and fighting. As many in-game cutscenes also feature steady dialogue scenes, we additionally produced a few calmer scenes, balanced between the directors. Each cutscene lasts thirty to sixty seconds, and the overall duration is the same for both directors.

In order to brief participants without pre-existing cinematic knowledge, we moreover prepared two mashup videos of real movie clips demonstrating the directors' iconic cinematographic style. These two reference videos have a length of two and a half minutes each. The reference and stimuli clips were embedded in an online form, starting with the reference clips but shuffled randomly within each group. The mashup videos and eight cutscenes can be viewed online\footnote{YouTube playlist: \url{https://youtube.com/playlist?list=PLm0M-oJsXvHpAu7dFCxTxcnoSVQI7ekzL}.}.

\subsection{Methods}
We employed a Binomial test to assess whether the generated cutscenes for one director were consistently perceived as different from the other's. The null-hypothesis is that our participants associate the stimuli cutscenes with one of the two directors by random chance, i.e. $p_0=0.5$. The Binomial test tells us whether this null hypothesis should be rejected, by measuring how significantly our participants' associations deviate from this hypothesised distribution. We performed the test on the pooled answers from all participants, which requires the repeated measures, i.e. associations, obtained from each participant to be independent. We consider this assumption valid as no feedback was given about the correctness of the associations, preventing any learning effects.

This null hypothesis would also be rejected if our participants associated the cutscenes consistently, i.e. beyond random chance, but with the \enquote{wrong} director. To validate the correct association of the generated cutscenes with the target director, we calculated the accuracy for each participant. %It measures the closeness of their individual measurements, i.e. associations, to the ground truth, i.e. the director used as input in the cutscene stimulus generation. The accuracy score requires the dataset to be balanced between the two directors, which is given. 

%In addition to these individual scores, w
We also calculated two inter-rater reliability measures to determine the conformity of our participants' individual associations \cite{GISEV2013330}. We firstly calculated Fleiss' Kappa \cite{fleiss1973equivalence}, an extension of Cohen's Kappa to more than two raters. We backed this up with a two-way mixed, average score ICC(3,k) Intraclass Correlation (ICC) analysis \cite{Shrout1979IntraclassCU}. This model requires that each subject is measured by $k$ fixed raters, and that the measures are averaged for each subject. This holds, as each cutscene was assessed by each participant, who were the only raters of interest.

We complement these quantitative measures with qualitative data from optional questionnaire responses and free-form discussion to learn about the participant's confidence in deciding on the directors, and how it was affected by our procedure. It comprised the following statements on 5-point Likert (Strongly Agree - Strongly Disagree) scales:
\begin{itemize}
    \item It was really hard for me to decide throughout the process.
    \item I felt like I was answering randomly.
    \item I needed longer clips/more time to accurately decide.
    \item The reference clips for the directors were helpful for me while deciding.
\end{itemize}
The questionnaire was provided as an online form and is included in our Supplementary Material.

\subsection{Participants}
We recruited volunteer adult participants via call-outs and advertisements by the research group through Telegram groups and Discord. Participants were deemed eligible if expressing some affiliation with movies and video games. %being at least 15 years old, and having
Exclusion criteria included lack of sleep, being under the influence of drugs or alcohol, as well as experiencing any digestive, muscle or organ pain, or emotional distress. Participants were not incentivised.

In total, 18 participants took part in the study. Three identified as female (16\%), 14 as male (77\%) and one chose to not provide demographics information. The reported age ranged from 22 to 41 years ($Median=27, SD = 5.85$). 18 participants are sufficient as we do not need to divide the population for comparing multiple experimental conditions.
%for hypothesis testing because we can pool their responses and thus obtain a large effective sample size.

% The missing male seems to be a problem on my part regarding the writing, I do not recall any participant leaving empty and reporting male afterwards, so I made a mistake while reporting that. The correction should be: 3 female (%16), 14 male (%77), 1 blank (%5). 

\subsection{Procedure}
The study was performed via individual online videoconferencing sessions between one participant and experimenter. We asked each participant to provide informed consent to participate in the study. %, and separate consent to watch reference clips and cutscenes with a potentially violent nature. 
They were then briefed about the \ProjectName{} system and its purpose. The goal of the next step was to familiarise the participant with the styles of our two directors. As preparation, we introduced them to basic camera manipulation techniques and the notion of directing style. The participants were then given access to an online form, in which they could watch the mashup clips for Quentin Tarantino and Guy Ritchie, respectively. They were asked to consider the main differences between shooting styles and camera management between the clips, and to pay attention to potential signature techniques. The participants were informed that they could watch these reference clips again at any time during the experiment. 

Once they were ready to proceed, our participants answered two optional demographics questions about their age and gender in the same form. This allowed them to rest their minds briefly before commencing with the main part of the experiment, for which they were asked to watch each stimulus cutscene and associate it with one of our two directors. 

The study concluded with a questionnaire and free-form discussion about our participants' decision-making, as described earlier. We provide the briefing material and the main experiment form with one randomisation of the reference and stimulus clips as Supplementary Material.

\subsection{Results}

\begin{table}[t!]
 \begin{tabular}{p {1.5cm} p{2.5cm} p{1.5cm} p{1.5cm}} 
 \textbf{Scene} & \textbf{Description} & \textbf{Director} & \textbf{Correct \%}\\
 \midrule
 Escape & Fight action & Tarantino & 88.8 \\ 
 \midrule
 Meeting & Thriller & Tarantino & 77.7 \\
 \midrule
  Facility & Spy action & Tarantino & 77.7 \\ 
 \midrule
  Village & Drama & Tarantino & 66.6 \\ 
 \midrule
  Bridge & Driving action & Ritchie & 94.4 \\
 \midrule
 Alley & Foot chase & Ritchie & 100 \\
 \midrule
 Night & Murder mystery & Ritchie & 72.2 \\ 
 \midrule
  Port & Drama & Ritchie & 55.5 \\
\end{tabular}
\vspace{0.3cm}
\caption{\label{table:cut}Cutscene information and percentage of correct predictions.}
\vspace{-2em}
\end{table}

Pooling the associations of 18 participants on eight stimuli each, we obtained an effective sample size of $n=144$. %The main part of our experiment yielded an effective sample size of $n=144$ associations, the pooled responses of 18 participants on eight stimuli each. 
The Binomial test on these measurements yielded $p=\num{9.93e-13}$ ($x = 114$) with a 95\% confidence interval $[0.716, 0.855]$, indicating that the null hypothesis of the associations being random should be rejected. This supports that \ProjectName{} can produce cutscenes that people consistently associate with a certain director beyond random chance.
%And a sample estimate of the probability of successfully guessing the ... of $p=0.7916667$.

The individual accuracy scores, reported in Table \ref{table:cut}, moreover support that the cutscenes were correctly associated with the target director. In total, 79\% of the 144 associations were correct. Further analysis of the individual scores highlights that some cutscenes were associated with considerably lower (\textit{Village}, \textit{Port}) or higher (\textit{Bridge}, \textit{Alley}) success than others. Strikingly, all 18 participants correctly recognised that the \textit{Alley} scene was shot with Guy Ritchie's data.

This is complemented with Fleiss $\kappa=0.382$ ($z=13.4, p=0$), indicating \textit{fair} inter-rater agreement \citep{landis1977measurement} between our participants. We moreover obtained an ICC value of $r=0.93$ ($p=\num{5e-13}$) with a 95\% confidence interval $[0.85, 0.98]$. Based on this upper and lower confidence bound, our ICC inter-rater reliability is between \textit{good} and \textit{excellent} \cite{TerryK}. 

The semi-structured interview revealed more nuanced information on our procedure's reliability. Most participants agreed to having experienced difficulty in associating some scenes due to their slow pace, as this was uncharacteristic of both directors. The general opinion was that the \textit{Port} scene was hardest to assess, \enquote{due to the lack of shooting and fighting}. The majority of participants agreed that the reference clips were extremely helpful for their decision-making, with the exception of three participants who criticised that the clips made it too easy to decide on a few cutscenes. Beyond the fixed questions, the interviews yielded that participants generally enjoyed the cutscenes generated for the study and found them suitable to tell a story. Most stated that they can spot the effect of the \ProjectName{} and how it tries to mimic a particular director. However, three participants mentioned that the absence of more realistic characters, lip-sync and eye animations, as well as a well-mixed sound design made it harder to associate the scenes, because they tried to infer the director based on characters, emotions and music, rather than through the camera manipulation.

\section{Experiment 2: Usability}
\label{section:usability}

While our first experiment was dedicated to testing \ProjectName{}'s functional value, we conducted this second experiment to probe the toolset's usability for game designers as required for its professional adoption. We wanted to (a) determine \ProjectName{}'s overall usability, (b) identify any major flaws affecting usability and user experience, and (c) learn how the toolset could be improved further. In individual videoconferencing sessions, we asked our participants to interact with \ProjectName{}, installed on their own computer, on a predefined sequence of tasks. In each of these single sessions, we assessed usability by means of standardised quantitative, and with qualitative methods. 

\subsection{Materials}
The main material for this usability study is the \ProjectName{} software. Implemented as a plug-in for the Unity game engine, its installation is straight-forward, but requires Unity, e.g. in form of the free Personal Edition, to be pre-installed. 

\begin{figure}[h!]
\centering
\includegraphics[width=0.6\columnwidth]{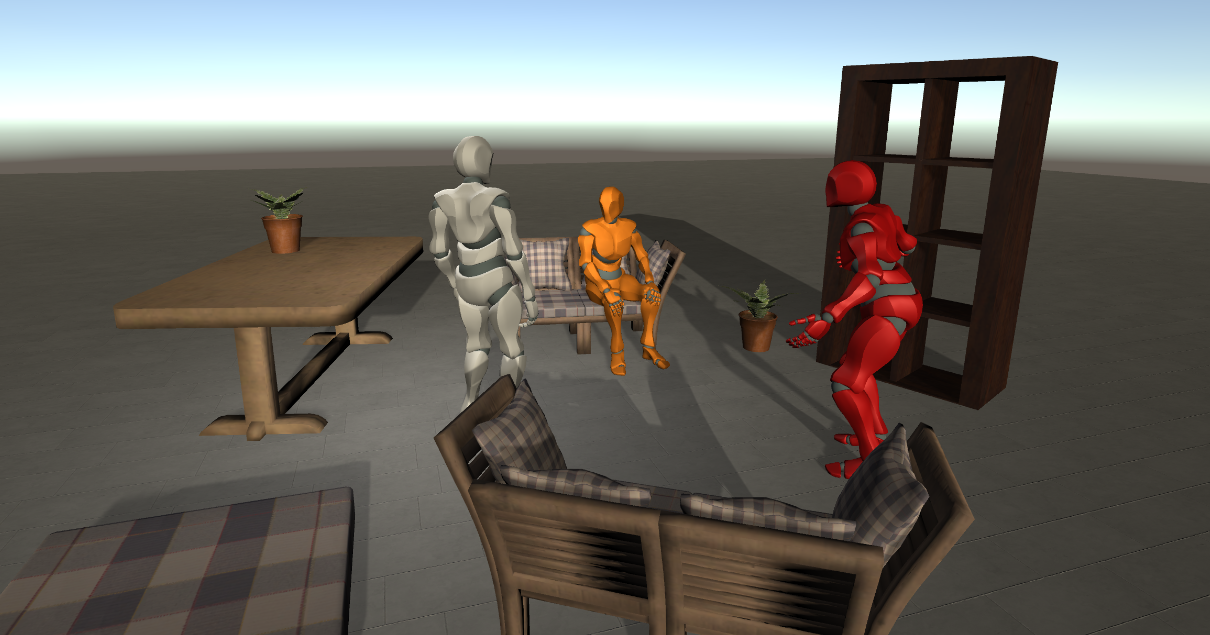}
\caption{The \enquote{Living room} 3D scene utilised in our usability experiment. Three characters, identified with grey, orange and red colors, have been animated to walk around and enact a conversation.}
\label{fig:livingroom}
%\vspace{-1.5em}
\end{figure}

We moreover developed a simple 3D scene (Figure \ref{fig:livingroom}, in the GitHub repository) to provide each participant with a consistent basis for their individual interaction with \ProjectName{}'s user interfaces. It includes various characters, engaged in a dialogue and walking around in a living room environment. To avoid distraction, the scene does not contain any gameplay elements, but only the cutscene animations ready to be played, with no camera motion attached to it.

\subsection{Methods}
We assessed usability qualitatively and quantitatively. We performed Concurrent Think-Alouds (CTAs) \citep{ericsson1984ha} to gain qualitative insights into our participants' thoughts while interacting with \ProjectName{}, potentially uncovering flaws and future improvements (b, c). This method has been well-vetted for the accurate collection of usability data \citep{Cooke}, and requires participants to verbalise their thoughts while performing a variety of tasks. 

We moreover employed the System Usability Scale (SUS) \citep{brooke,jR} to obtain a standardised, quantitative measure of \ProjectName{}'s usability (a). Initially developed as a quick means to assess software usability -- comprising only ten items on 5-point Likert (Strongly Agree - Strongly Disagree) scales -- it has proven a reliable instrument for the end-of-test subjective assessments of usability, yielding an easily interpretable score from 0 to 100\% \cite{jR}.

We finally conducted a a semi-structured interview to collect more targeted, qualitative information on flaws (b) and potential improvements (c) in the system's usability, and to receive feedback on the study and procedure.

\subsection{Participants}
Since \ProjectName{} is implemented in Unity, and we wanted to keep interference with general, experience-related  usability issues minimal, we required our participants to have prior practice in using the game engine. They had to be capable of navigating the Unity user interface, using the timeline tool, and test their work both in the editor and game modes. To accommodate this requirement, we recruited our study participants through the Unity online forum and a Telegram group for [removed for anonymisation] Game \& Media students. The eligibility and exclusion criteria were otherwise identical to our first experiment. We did not provide incentives for participation.

In total, twelve adult participants took part in the experiment, of which three identified as female (25\%) and nine as male (75\%). Participants included professional animators, 3D artists, and Unity developers. We consider this sufficient, since we do not compare experimental conditions and complement usability scores with qualitative data.  

\subsection{Procedure}
The study was performed in individual videoconferencing sessions between one participant and the experimenter. The participant was provided the \ProjectName{} plug-in for installation with their local copy of Unity ahead of the study session (\url{https://github.com/inanevin/Cine-AI}). In the beginning of the live session, they received a briefing on the capabilities and purpose of the \ProjectName{} toolset (Supplementary Material). As a consistent primer to the toolset, they were shown the windows and asset files required to set-up and use the system.

For the main part of the study, the participants were asked to think aloud while performing various tasks, thus performing a CTA. The tasks consisted of the initial system set-up, importing the sample director data, defining scene proxies, adding shot markers in the timeline, running the simulation, and tweaking the results through the storyboard interface. The experimenter took notes of any thoughts and comments related to \ProjectName{}'s usability.%-We presented this scene to the users and asked them to start using \ProjectName{}, adding shot markers and simulating them in order to produce a version of the cutscene usable in a game.

Upon completion of the demo scene simulation, participants were asked to fill in the SUS questionnaire and provide optional demographics information on their age and gender. This was followed by the semi-structured interview to identify additional usability flaws and means of improvement. The main questionnaire with the demographics and SUS items, as well as the optional, semi-structured interview questions are comprised in the Supplementary Material.

\subsection{Results}

Across all participants, \ProjectName{} received a mean SUS score of 74.36\% ($Median = 77.5, SD = 17.50$). Compared against a reference dataset \citep{Sauro2011} from 446 studies and over 5000 individual SUS responses ($Mean=68,  SD=12.50$), this corresponds to a percentile rank of 70\% \cite{SAURO2012185}, indicating that \ProjectName{}'s usability is above average and better than seven out of ten reference systems. On a curved grading scale from A to F \cite{SAURO2012185}, \ProjectName{} receives a usability grade B.

The CTA did not uncover any major usability issues; all participants were able to work through the given tasks without severe problems. Regarding minor issues, some participants lamented the lack of tooltips with more information on the adjustable parameters. The storyboard interface was generally well received, with participants highlighting its professional look and usefulness in visualising their composition. Some participants suggested that the storyboard could be improved by offering better layout options and the ability to use multiple aspect ratios for shot previews. Three participants complained about the set-up procedure of the toolset and one requested a wizard to automate it further. Additional feature requests concerned \ProjectName{}'s embedding into the Unity game engine. Two participants expressed a preference for completely integrating the toolset within Unity's own tools, in particular runtime editing support and animation baking in Unity's timeline. 

\section{Discussion}

Our first experiment (Section \ref{section:study_comparison}) demonstrates \ProjectName{}'s capability to generate in-game cutscenes in a target director’s style that is recognisable and distinguishable by its users. The participants in our second experiment (Section \ref{section:usability}) did not identify any fundamental flaws to the system's usability, and indicated that \ProjectName{} has potential to be adopted in professional cutscene production. 
%provides evidence that \ProjectName{} complements this core functionality with above-average usability, a key requirement for the system’s adoption in actual projects. 
We next discuss present and permanent constraints to  \ProjectName{}'s functionality and usability, critically reflect on limitations to our experiments, and propose how these can be overcome in future work.

\subsection{Director Style Imitation}
%Barely any directors have perfectly distinct styles, since 
%Looking at movie directors at large, it is extremely difficult to identify an individual style that is perfectly distinct from anyone else's.
%We challenge our reader to find a director whose style is perfectly distinct from anyone else's. 
For any selection of directors, especially if working in a similar genre, individual style can rarely be perfectly distinguished. This is because style evolves over time and typically manifests only in specific types of scenes. The latter observation can explain the outliers in our first experiment: our participants likely experienced strong difficulty in correctly associating the \textit{Village} and \textit{Port} scenes with their target directors, because %the scenes showed strong stylistic overlap:  
%to demonstrate their iconic styles, these cutscenes contained less iconic cinematography styles but more common ones. One can assume that the success rate will drop as the genre of the cutscenes move further away from the sample directors' preferred styles. 
both scenes contained more dialogue and less action, while our directors' iconic style is mostly expressed in action and fast-paced content. We explain our participants' strong success rate in correctly associating the \textit{Alley} scene with Guy Ritchie by the presence of a chase sequence that was extremely similar to a scene in Ritchie's reference clips. These properties of style impose fundamental limitations to the accuracy by which \ProjectName{} or any similar system could imitate a particular director.
% Alley scene: thus overfitting?
% Subjectivity: would be good to back this up with a second study in which original clips are replicated in the Unity Scenes, thus CineAI not involved. Is the association accuracy still imperfect?
% Apart from stylistic overlap, the association of a cutscene with a specific director is also shaped by subjectivity, as highlighted by our two inter-rater reliability measurements. We hence consider the identified accuracy a sufficiently strong result to underline this aspect of \ProjectName{} functionality.

We also see room for improvement, based on the observation that not all directors solely focus on camera work to express their style. Michael Bay for instance notoriously employs many post-processing effects such as lens flares and god rays. To increase \ProjectName{}'s fidelity in distinguishing and reproducing a director's unique style, we thus suggest to dedicate some future work to implementing and annotating additional cinematography techniques of different types, including post-processing effects. 
%Tim Burton, a well-accomplished director known for his Gothic style, achieves his unique results mostly through dark and edgy character design \cite{burton2008burton}, and 
%  While character design is out of scope for \ProjectName{}, it can in principle be extended to apply post-processing effects to mimic the style of Michael Bay and others.
Moreover, we recommend to encode and interpret more information about the ground truth clips beyond the use of specific techniques, e.g. concerning the characters and scene objects, to additionally boost \ProjectName{} imitation capacity. While such additional features only have to be implemented once, they crucially require more review and annotation work in the creation of every new director imitation dataset.
%Such cases might occur, where the directors tend to use common and non-distinguishable camera techniques but accomplish their signatures through other means. In such cases, it would not be reasonable to expect \ProjectName{} to work accurately. In future work, it is possible to incorporate more aspects of filmmaking into \ProjectName{}, such as more information about the characters and entities in the cutscene, post editing features and visual effects. By encoding director data for other aspects of filmmaking and implementing these aspects in \ProjectName{}, it is possible to significantly increase the accuracy of the imitation process.

We consider our style imitation experiment to be presently limited by the choice of Quentin Tarantino and Guy Ritchie as target directors. 
%Our style imitation experiment has been limited by our choice to imitate the style of Quentin Tarantino and Guy Ritchie. 
While allowing for a solid proof-of-concept due to the characteristics discussed in Section \ref{section:data_director_selection}, both directors focus on the action genre and exhibit distinctive and iconic styles that can be recognised with only moderate difficulty. To better assess \ProjectName{}'s fidelity and universal applicability, future research must evaluate its capacity to imitate additional directors working in other genres and with less clearly distinguishable styles. 
% We consider a particular limitation of the present study that the the selection of ground-truth clips and their analysis is biased by the authors' specific cinematography knowledge. 

Both directors are moreover white and western men, and future efforts in extending the dataset should focus on increasing director diversity. We would be intrigued to see how well \ProjectName{} can for instance reproduce the style of the Indonesian director Timo Tjahjanto, who works on horror and action movies and employs superior camera work in extremely fast-paced scenes without creating visual discomfort. The American  director Spike Lee focuses on color and race relations and is well known for his frequent use of dolly shots to let characters \enquote{float} through their surroundings. Finally, the Belgian-born French director Agnes Varda has been highly praised for her unique style of using the camera \enquote{as a pen}; this provides a worthwhile challenge to procedural cinematography.

We finally note that our dataset is limited by the selection of ground-truth clips and their annotation being dependent on the authors' cinematography knowledge. The dataset creation process affects \ProjectName{}'s performance and, in comparison against human judgement, its imitation accuracy scores. As a straight-forward means to reduce subjective bias in the future, we propose employing more annotators, and validating the reliability of the resulting dataset with inter-rater agreement measures. As a more ambitious plan, we envision to overcome such bias, as well as avoiding an increase in labour due to more features, by automating the annotation process via machine learning. This is ambitious, in that the respective algorithms must be capable of differentiating cinematography techniques from image sequences, which requires advances to the current state of the art. %Such automation would allow to rapidly produce distinct director data and categorize them into different use-cases for possible game scenarios. Having multitude of director data samples, \ProjectName{} would be able to produce at least initial results for scene directing and might remove the need of hiring an actual director.

\subsection{Usability}
Our usability study is limited in that participants were subjected to a \ProjectName{} version that was not tightly integrated into the Unity game engine and editor, potentially impeding usability. While this decoupling was intentional to allow for the straight-forward migration to other systems such as the Unreal Engine, we aim to accommodate our participant's engine-specific feature requests and leverage more intuitive and accessible native UI components in future work.

%We moreover propose to improve \ProjectName{}'s usability through modifications to the timeline and storyboard, our two core UI components. At present, a simulation of the timeline markers yields a linear scene composition with one transition between any two markers; if a user wants to change this composition, they presently have to run the simulation again. For future versions, we aim to realise multiple transitions between the markers, thus presenting the user with multiple scene compositions to choose from without the need for repeated simulation.

We moreover propose to improve \ProjectName{}'s usability through modifications to the timeline and storyboard, our two core UI components. At present, Cine-AI simulates all markers once and presents them in the storyboard, through which the user can re-simulate all or individual markers for fine-tuning. In future versions, we want to generate multiple options for each marker for the user to select from in the storyboard, thus alleviating effort in re-simulating markers.% and present them in storyboard. Thus, the user would be presented with variety of options for each marker, which they can select from. This would lighten the load by lowering the need to re-simulate markers.

%In order to not overburden the user with choosing a particular composition, we plan to rank them based on the \textit{screenplay} \citet{lino2011director} of any marker within, a data structure containing information about a shot, including the scene information, involved actors and their 3D object data, as well as the textual descriptions of the actions performed by the actors.

%Therefore, it would be possible to provide various transition and shot suggestions to the users. \citet{lino2011director} introduce a ranking system for possible shot compositions, based on the notion of a \textit{screenplay}. Screenplay is a data structure containing information about a shot, including the scene information, involved actors and their 3D object data, as well as the textual descriptions of the actions performed by the actors. The screenplay data is used to compute suggestions for automated camera motion. We plan to implement a similar system, where our system will generate multiple shots based on different cinematography techniques and rank them by using the previous shot marker's screenplay data. Therefore, it will be possible to create a more professional toolset ready for advanced production pipelines.

We plan to improve the usability and functionality of our storyboard by allowing users to define their own cinematography rules based on existing, specialised declarative languages \citep{ChristiansonDeclarative}. Combined with a constraint solver to prevent rule conflicts, such user-defined rulesets would complement presently hard-coded rules such as \citet{Arijon}'s idioms (Sec.~\ref{sec:system_selecting}), increasing \ProjectName{}'s expressive power and customisability.

\subsection{Performance and Automation}
We highlight two avenues for increasing \ProjectName{}'s performance and automation, drawing on related work (Sec.~\ref{sec:related_work}). In order to achieve faster and more accurate camera placement, future work should be dedicated to implementing \citeauthor{linoref2}'s Toric Space \cite{linoref2}. This novel camera space representation uses a triplet of Euler angles to define a camera viewpoint around a pair of targets, enabling automated viewport computation. %The unfolding calculation presents a set of manifolds, which contain data about possible camera placements and orientation angles that ensures subject visibility. 
It represents a fast, uniform way of determining possible camera placements given any target in 3D space. At present, each cinematography technique in \ProjectName{} must define its own camera placement rules. We suggest to overcome this by integrating Toric Space in \ProjectName{}, and only exposing the calculation parameters. This would make it considerably easier to include new cinematography techniques: a new instance would be defined by a set of manipulation parameters alone.

The use of Toric Space opens up a second avenue for future work. \ProjectName{}'s scene proxies can be slow to compute in large, complex 3d scenes and on lower-end hardware as they rely on checking all collision information within the volume and serializing the data for cinematography technique calculations. One option for future work is to turn these into runtime calculations with an efficient and precise algorithm such as \citeauthor{burg}'s occlusion maps \cite{burg}, which relies on \citeauthor{linoref2}'s Toric Space \cite{linoref2}. Such an algorithm would ensure high-rates of subject visibility while obeying the cinematography rules. Moreover, it would eliminate the need to improve the present runtime camera helpers, thus providing better responses in fully dynamic scenarios.

\subsection{System Adoption}
We hope that \ProjectName{}'s capacity to imitate a given director's style, its usability, and its implementation in the popular and freely available game engine Unity will foster adoption in real-life game projects. Crucially, we do not expect automation through \ProjectName{} to take away jobs in game industry. We consider our research and the developed tool to be particularly valuable for independent game developers, who can typically not afford a dedicated production team for creating in-game cinematography in the first place. We do not expect \ProjectName{} to completely replace the process of manual directing, but rather see its potential to co-create an initial cutscene in the consistent and uniform style of a certain director without specialised cinematography expertise. The result can then be tweaked further by the independent game designer to fit their game's unique style. The cinematography requirements of AAA games tend to get extremely sophisticated in terms of style, scene duration and size. While AAA companies will thus likely continue relying on dedicated production teams to achieve the desired level of cinematographic quality, they can utilise \ProjectName{} to create prototypes for their cinematography design, automatically generate shots to inspire new ideas, or use the storyboard feature to quickly iterate on possible shots, following a specific directorial style.
%We expect this to be especially appealing to independent game developers, who often lack the resources to hire dedicated cutscene production teams. Across different applications, \ProjectName{} could handle most of the ground work by automating the camera management process.
%It is not likely for most independent developers to afford a dedicated production team merely for creating in-game cinematics. Cine-AI can help tremendously in such cases, by automatically generating the baseline for a given cut-scene. Independent developers can then extend the procedurally generated shots to complete their cut-scenes, saving a great amount of time and budget.
%Especially independent developers could use \ProjectName{} to get ideas about how their favourite director would shoot their cutscene. 
%We expect this to be especially appealing to independent game developers, who often lack the resources to hire dedicated cutscene production teams. Across different applications, \ProjectName{} could handle most of the ground work by automating the camera management process.

\section{Conclusion}
We have presented \ProjectName{}, a novel, semi-automated cinematography toolset capable of procedurally generating in-game cutscenes in the style of a specific movie director. The system combines a highly configurable design-time workflow with runtime support, allowing for cutscenes to unfold dynamically based on the game state. Through extensive tools, including a novel storyboard interface, the user can adjust simulation properties and tweak the final scene composition. Via two user studies, we provided evidence that \ProjectName{} can generate cutscenes that people correctly associate with a specific target director, and that the system provides above-average usability. We point out present limitations of our dataset and study, and discuss potential means of adopting \ProjectName{} for production in game industry studios of different sizes. We highlight opportunities for future work to improve style imitation fidelity, increase design- and runtime efficiency, foster extensibility, and improve automation further. 

To support \ProjectName{}'s adoption and future extension, we made our proof-of-concept dataset and the source code publicly available under an open-source license. We invite researchers, (game) developers and film enthusiasts to join us in taking this project further.

\section{Acknowledgments}\label{sec:acknowledge}
We thank our reviewers for their thorough and constructive comments. CG was partly funded by the Academy of Finland Flagship programme ``Finnish Center for Artificial Intelligence'' (FCAI). 

\bibliographystyle{ACM-Reference-Format}
\bibliography{main}

\received{February 2022}
\received[revised]{June 2022}
\received[accepted]{July 2022}

\end{document}